\documentclass{nature}
\usepackage{aas_macros}

\usepackage{color}
\newcommand{\revised}[1]{\color{black}#1}
\newcommand{\revisedtwo}[1]{\color{black}#1}

\title{The origin of polarization in kilonovae and the case of the gravitational-wave counterpart AT\,2017gfo}

\author{M. Bulla$^{\ref{OskarKlein}}$, S. Covino$^{\ref{Brera}}$, K. Kyutoku$^{\ref{KEK},\ref{Sokendai},\ref{iTHES},\ref{KyotoUniv}}$, M. Tanaka$^{\ref{Tohoku},\ref{Tokyo}}$, J.~R. Maund$^{\ref{Sheffield}}$, F. Patat$^{\ref{ESO}}$, K. Toma$^{\ref{Tohoku},\ref{SendaiFrontier}}$, K. Wiersema$^{\ref{Leicester},\ref{Warwick}}$, J. Bruten$^{\ref{Sheffield}}$, Z. P. Jin$^{\ref{PurpleMount}}$ \& V. Testa$^{\ref{MPorzio}}$}

\begin{document}

\maketitle

\begin{affiliations}
 \item 
 \label{OskarKlein}
 Oskar Klein Centre, Department of Physics, Stockholm University, SE-106 91, Stockholm, Sweden
 \item
\label{Brera}
Istituto Nazionale di Astrofisica / Brera Astronomical Observatory, via Bianchi 46, 23807 Merate (LC), Italy.
\item
\label{KEK}
Theory Center, Institute of Particle and Nuclear Studies, KEK, Tsukuba 305-0801, Japan
\item
\label{Sokendai}
Department of Particle and Nuclear Physics, the Graduate University for Advanced Studies (Sokendai), Tsukuba 305-0801, Japan
\item
\label{iTHES}
Interdisciplinary Theoretical and Mathematical Sciences Program (iTHEMS), RIKEN, Wako, Saitama 351-0198, Japan
\item
\label{KyotoUniv}
Center for Gravitational Physics, Yukawa Institute for Theoretical Physics, Kyoto University, Kyoto 606-8502, Japan
\item
\label{Tohoku}
Astronomical Institute, Tohoku University, Aoba, Sendai 980-8578, Japan
\item
\label{Tokyo}
National Astronomical Observatory of Japan, National Institutes of Natural Sciences, Osawa, Mitaka, Tokyo 181-8588, Japan
\item
\label{Sheffield}
Department of Physics and Astronomy, University of Sheffield, Hicks Building, Hounsfield Road, Sheffield S3 7RH, UK
\item
\label{ESO}
European Southern Observatory, Karl-Schwarzschild-Str. 2, D-85748 Garching bei M\"unchen, Germany
\item
\label{SendaiFrontier}
Frontier Research Institute for Interdisciplinary Sciences, Tohoku University, Sendai 980-8578, Japan
\item
\label{Leicester}
Department of Physics \& Astronomy and Leicester Institute of Space \& Earth Observation, University of Leicester, University Road, Leicester LE1 7RH, UK
\item
\label{Warwick}
University of Warwick, Coventry, CV4 7AL, UK
\item 
\label{PurpleMount}
Key Laboratory of Dark Matter and Space Astronomy, Purple Mountain Observatory, Chinese Academy of Sciences, 210008, Nanjing, China
\item
\label{MPorzio}
Istituto Nazionale di Astrofisica / Osservatorio Astronomico di Roma, via Frascati 33, 00078 Monte Porzio Catone, Italy 
\end{affiliations}

\begin{abstract}
\revisedtwo{The Gravitational Wave (GW) event GW\,170817 was generated by the coalescence of two neutron stars (NS) and produced an electromagnetic  transient, labelled AT\,2017gfo, that was target of a massive observational campaign. Polarimetry, a powerful diagnostic tool for probing the geometry and emission processes of unresolved sources, was obtained for this event. The observed linear polarization was consistent with being mostly induced by intervening dust, suggesting that the intrinsic emission was weakly polarized ($P < 0.4-0.5$\%). In this paper, we present and discuss a detailed analysis of the linear polarization expected from a merging NS binary system by means of 3D Monte Carlo radiative transfer simulations assuming a range of possible configurations, wavelengths, epochs and viewing angles. We find that polarization originates from the non-homogeneous opacity distribution within the ejecta and can reach levels of $P\sim1$\% at early times (1$-$2\,days after the merger) and in the optical R band. Smaller polarization signals are expected at later epochs and/or different wavelengths. From the viewing-angle dependence of the polarimetric signal, we constrain the observer orientation of AT\,2017gfo within $\sim$~65$^\circ$ from the polar direction. The detection of non-zero polarization in future events will unambiguously reveal the presence of a lanthanide-free ejecta component and unveil its spatial and angular distribution.}
\end{abstract}


The discovery of GW\,170817\cite{Abbottetal2017a} and its electromagnetic counterpart AT\,2017gfo has definitely been an epochal event. It was generated by the merging of two NSs and produced a transient electromagnetic source, dubbed ``kilonova" or ``macronova" (hereafter referred to as kilonova). This transient event was intensively followed with all the main ground-based and space-borne facilities\cite{Abbottetal2017b}, thus allowing us to study the evolution of the kilonova and later of the afterglow of the Gamma-Ray Burst GRB 170817A\revised{\cite{Mooleyetal2018b,Ghirlandaetal2018}.}
The optical/near-infrared observations were carried out by several teams\cite{Covinoetal2017,Evansetal2017,Pianetal2017,Smarttetal2017,Tanviretal2017} and delivered an almost continuous spectro-photometric and polarimetric coverage of the kilonova for about a couple of weeks since discovery. Some of the main observables for this category of sources were predicted a long time ago and the general agreement with the observational results is truly remarkable\cite{Li&Pacynski1998,Metzgeretal2010,Roberts2011,Kasenetal2013,Barnes&Kasen2013,Tanaka&Hotokezaka2013,Baiotti&Rezzolla2017,Metzger2017,Tanakaetal2017,Tanakaetal2018}. Nevertheless, several details -- like ejecta mass, velocity, composition and distribution -- are still unclear despite them being crucial to e.g. turn the GW detection into true rates, allow a comparison of kilonova detection limits from optical and infrared surveys with GW data and compare the heavy-element yields with cosmic abundances. It is therefore of paramount importance to identify new observational diagnostics able to give complementary information. 

Among these, optical and near-infrared polarimetry is of great interest since it can be sensitive to the intrinsic geometry of the source, the composition of the emitting region material, and their dynamical evolution. Some of these properties are not easily constrained by the analysis of (current) ordinary light curves and spectra. On the other hand, apart from a few general considerations\cite{2013PhRvD..88d1503K,Kyutokuetal2015}, there are no detailed and quantitative published studies describing the expected linear polarization signatures from a kilonova produced by the merging of a NS binary system. In this paper, we first discuss the theoretical scenario adopted to model the kilonova evolution from about 1\,day to roughly a week after the GW event. We then present linear polarization predictions for different wavelengths and observer orientations, and discuss the role played by several factors that cannot be unambiguously determined from other observations, such as the geometry and the opening angles of the adopted ejecta components. Finally, we provide some guidelines to drive future polarimetric observations of kilonovae, with or without accompanying GW data. A more detailed description of the adopted models, radiative transfer simulations and results is given in the Methods.

\section*{Kilonova models}

Both the outcomes of simulations\cite{Hotokezaka2013,Bauswein2013,Fernandez2013,Siegel2017,Shibata2017,Fujibayashi2017,Metzger2014,Lippuner2017} and the analysis of the spectro-photometric observations of AT\,2017gfo\cite{Pianetal2017,Smarttetal2017} suggest that kilonovae can be interpreted as the combination of two ejecta constituents: a first component distributed around the equatorial plane, characterized by high opacities of lanthanide elements (\revised{lanthanides are a subclass of $r-$process elements characterized by atomic number $57\leq Z\leq71$}) and giving rise to a ``faint-and-red" kilonova, and a second component in the polar regions, characterized by relatively lower opacities and producing a ``bright-and-blue" kilonova. Motivated by these considerations, we construct a two-component ejecta model where (i) high-opacity lanthanide-rich materials are distributed close to the equatorial plane while (ii) lanthanide-free materials with lower opacities occupy regions at higher latitudes (see Fig.~\ref{fig:sketch}). Despite the general agreement between the two-component model and AT\,2017gfo light curves and spectra, we note that current models\cite{Hotokezaka2013,Fernandez2013,Perego2014,Fujibayashi2017} struggle to reproduce the high velocities of the blue component and the high masses of the red component inferred for AT\,2017gfo\cite{Chornock2017,Cowperthwaite2017,Nicholl2017,Pianetal2017,Smarttetal2017}. However, the general polarization behaviours predicted in this work (such as the viewing angle, wavelength and time dependence of the signal) are expected to hold with using a different and/or more sophisticated kilonova scenario.

We construct our fiducial model following suggestions from both hydrodynamical simulations and the analysis of the spectro-photometric data of AT\,2017gfo\cite{Hotokezaka2013,Bauswein2013,Fernandez2013,Siegel2017,Shibata2017,Fujibayashi2017,Metzger2014,Lippuner2017,Pianetal2017,Smarttetal2017,Chornock2017,Cowperthwaite2017,Nicholl2017}. Specifically, we adopt \revised{spherical ejecta with} an half-opening angle of $\Phi=30^\circ$ for the lanthanide-rich region, set the whole ejecta mass between $v_\mathrm{in}=0.05$~c and $v_\mathrm{out}=0.3$~c to $M_\mathrm{ej}=0.03\,M_\odot$  and use a power-law density profile $\rho(v)=\rho_\mathrm{in}\,(v/v_\mathrm{in})^{-3}$, where the density at $v_\mathrm{in}$ is $\rho_\mathrm{in}=1.2\times10^{-12}$~g~cm$^{-3}$. We place the photospheric surface at $v_\mathrm{ph}=0.15$~c, which at 1.5~days from the merger corresponds to a photospheric density of $\rho_\mathrm{ph}=5.6\times10^{-14}$~g~cm$^{-3}$ and $\sim0.01\,M_\odot$ of ejecta mass above the photosphere. \revisedtwo{The ejecta are assumed to be in homologous expansion (see Methods - Our model: densities, opacities, and velocities.).} Using the Monte Carlo radiative transfer code \textsc{possis}\cite{bulla2015,bulla2017}, we calculate polarization levels for our fiducial model at different epochs and wavelengths. Photon packets are created unpolarized at $v_\mathrm{ph}$, emitted assuming constant surface brightness and propagated throughout the whole ejecta where they can be either polarized by Thomson (electron) scattering or depolarized by bound-bound line interactions. Electron scattering and wavelength-dependent bound-bound opacities are taken from numerical simulations\cite{Tanaka&Hotokezaka2013,Tanakaetal2018}, and their time-dependence taken into account to investigate the polarization at different epochs (see Methods - Opacity calculations for lanthanide-rich and lanthanide-free ejecta \revisedtwo{for further details}). Photon packets escaping the ejecta are collected and used to calculate continuum polarization signals at different viewing angles $\theta_\mathrm{obs}$, where $\cos\theta_\mathrm{obs}=0$ corresponds to an observer in the equatorial plane. 

\section*{Polarization}

\revised{Results of the polarimetric campaign devoted to AT\,2017gfo are reported in \cite{Covinoetal2017}. Observations at five epochs, from about 1.5\,days to almost 10\,days, were secured. A linear polarization of $P = (0.50 \pm 0.07)$\% was measured during the first epoch, while stringent upper limits were placed on the following epochs, all consistent with the former measurement. The observed polarization level was consistent with that shown by several stars in the field of view of the optical counterpart, and therefore could be totally due to polarizing effect of dust along the line of sight\cite{Serkowskietal1975}.} 

In contrast to the case of supernovae (where the overall continuum polarization signal is driven by the shape of the photosphere\cite{Wang&Wheeler2008}), we find that the origin of linear polarization in kilonovae mainly resides in the asymmetric distribution of lanthanide-rich material in the ejecta, resulting in higher line opacities in regions closer to the equatorial plane. This leads to a net polarization signal since radiation from low latitudes is typically depolarized by line interactions while that from higher latitudes is more likely polarized by electron scattering interactions (see inset of Fig.~\ref{fig:pol}) \revised{at favourable wavelengths and times (see below)}. The polarization levels predicted at rest-frame wavelengths (7000~\AA, R band) and epochs (1.5~days) corresponding to the first polarimetric observation of AT\,2017gfo are shown in Fig.~\ref{fig:pol}. \revised{Owing to the axial symmetry of the model, the linear polarization signal is carried by the Stokes parameter $Q$ while the Stokes parameter $U$ is null (see caption of Fig.~\ref{fig:pol} for a definition of $Q$ and $U$), i.e. $P \equiv \sqrt{Q^2+U^2} = |Q|$.} The polarization signal peaks at $P$~\revised{$=|Q|$} $\sim0.8\%$ for an observer in the equatorial plane ($\cos\theta_\mathrm{obs}=0$) and then decreases to smaller and smaller values moving towards the pole ($\cos\theta_\mathrm{obs}=1$) since both \revised{ejecta components are axially symmetric by construction and thus} become increasingly closer to circular symmetry in projection. Our simulations show that signals of the order of $1\%$  may therefore be detected for future kilonova events observed \revised{in the R band}, at early times and at favourable viewing angles (i.e. observer orientations close to the equatorial plane). Such polarization levels are detectable for a wide range of magnitudes with current instrumentation.

Owing to the strong time and frequency dependence of line opacities, the polarization signal is found to vary considerably as a function of both epoch and wavelength (see Fig.~\ref{fig:timewave}). First, we predict a drop in polarization on short time-scales following the rapid increase of line opacities with time in both lanthanide-free and lanthanide-rich ejecta (due to a rapid temperature drop and neutralization, \revised{see Figs.~\ref{fig:opac}-\ref{fig:opac_time_wave}}). At 7000~\AA, the polarization level along the equator decreases from $\sim0.8\%$ at 1.5~days (first epoch of AT\,2017gfo) to only $\sim0.1\%$ one day after (second epoch), becoming negligible at later epochs \revised{(see Fig.~\ref{fig:timewave})}. Secondly, the strong wavelength-dependence of line opacities and the presence of two different ejecta components combine to give a maximum polarization signal around 7000~\AA. Moving to shorter wavelengths (5000~\AA), depolarizing bound-bound transitions become the dominant source of opacities in both ejecta components \revised{(see Fig.~\ref{fig:opac})} and thus the resulting polarization level is smaller ($P\lesssim0.1\%$ at all viewing angles). At longer wavelengths (10\,000 and 15\,000~\AA), instead, \revised{the relative importance of electron scattering to the total opacity increases in the lanthanide-rich ejecta. Compared to the case at 7000~\AA, some polarizing contributions from the lanthanide-rich ejecta cancel part of the polarizing contributions from the lanthanide-free ejecta , thus biasing the overall polarization signal to smaller values.} As a result, our simulations clearly indicate early-time spectropolarimetry in the R band as the best observational strategy to detect polarization in future kilonova events. 

\section*{Polarization dependence on model parameters}

Absolute polarization levels depend on three main parameters that are uncertain and difficult to estimate from both hydrodynamical simulations and the analysis of the spectro-photometric observations of AT\,2017gfo: the overall shape of the ejecta, the half-opening angle $\Phi$ and the photospheric velocity $v_\mathrm{ph}$. 

The polarization signal imprinted by the global shape of the ejecta is subdominant compared to that induced by the asymmetric distribution of lanthanide-rich material. \revised{Our calculations (see Methods - Simulation results) indicate that} ellipsoidal ejecta with axial ratio of 0.75 (i.e. semi-minor axis 25~per~cent shorter than semi-major axis) and lanthanide-free opacities would produce only $P\lesssim0.04\%$, a level which is much lower than those obtained from the asymmetric distribution of lanthanide-rich materials ($P\sim0.8\%$, see Fig.~\ref{fig:pol} and discussion above). 

The impact of $\Phi$ on the polarization signal is illustrated in Fig.~\ref{fig:opang}. For observers close to the equatorial plane ($\cos\theta_\mathrm{obs}\lesssim0.2$), the maximum polarization signal is reached for $\Phi=30^\circ$ when the solid angle subtended by the lanthanide-rich ejecta component is $2\pi$, i.e. half of the full solid angle (see inset in Fig.~\ref{fig:pol}). The partial coverage of the photosphere by the lanthanide-rich ejecta is less effective in polarizing the escaping flux for both lower and higher values of $\Phi$, therefore resulting in lower signals. Nevertheless, similar polarization degrees are found for $\cos\theta_\mathrm{obs}\gtrsim0.4$ regardless of the value of $\Phi$. \revised{For favourable viewing angles,} comparisons between our models and the combination of future spectro-photometric and polarimetric observations of kilonovae will be critical to constrain the angular structures of the two ejecta components, a result which can not be achieved by any other means.

The exact location of the photosphere (expressed in terms of the photospheric velocity $v_\mathrm{ph}$) has a relatively large impact on the polarization signal. In general, moving the photosphere deeper inside the ejecta leads to larger polarization since less unpolarized radiation manages to escape the lanthanide-rich regions. For instance, placing the photosphere at $v_\mathrm{ph}=0.1$~c gives a polarization of $P\sim1.9\%$ for an equatorial viewing angle, roughly twice as large compared to the fiducial case with $v_\mathrm{ph}=0.15$~c. In addition to the uncertain values of $v_\mathrm{ph}$, we note that the absolute values of polarization might also be dependent on our assumption of a single photosphere. In reality, the different opacities in the two ejecta components correspond to different locations of the photosphere, with the lanthanide-rich region producing photons farther out compared to the lanthanide-free region. 
We defer a more detailed treatment of the photosphere to a future study (when new hydrodynamical models and observational data will likely clarify the scenario) while stressing that the general polarization behaviours identified in this work are expected to hold.


\section*{Constraints on the polarization of AT\,2017gfo}

Beside predicting polarization signatures for kilonovae, our simulations are also crucial to estimate the intrinsic signal in AT\,2017gfo (and future events). As shown in Fig.~\ref{fig:timewave}, the R band polarization level turns out to be negligible at all viewing angles starting from $\sim$~2 days after the merger. The signal detected after $\sim$~2 days in AT\,2017gfo\revised{\cite{Covinoetal2017}} is therefore dominated by interstellar polarization (ISP)\revised{, which we estimate to be $P_{\rm dust} = (0.49 \pm 0.05)$\%.} Reasonably assuming that the ISP is the only source of polarization at late epochs and subtracting it from the one at 1.5 days \revisedtwo{(see Methods - Upper limits to AT 2017gfo intrinsic polarization)}, a common procedure in supernova polarimetry\cite{Maund2007,Wang&Wheeler2008,Patatetal2012}, leads to an upper limit on the intrinsic polarization in AT\,2017gfo of $P < 0.18$\% at 95\% confidence level. From the comparison between this upper limit and our model predictions in Fig.~\ref{fig:pol}, we conclude that AT\,2017gfo was observed within 65 degrees from the polar axis ($\cos\theta_\mathrm{obs}\gtrsim0.4$), \revised{a constraint that is less restrictive but consistent} with previous estimates based on spectral modelling and analysis of the associate short GRB\,170817A\cite{troja2017,Pianetal2017,Finstadetal2018} or on the use of independent distance measurements to break the inclination/distance degeneracy of the GW signal\cite{Abbottetal2017a,Mandel2018}. From Fig.~\ref{fig:opang}, we note that the constraints on the inclination of AT\,2017gfo are rather insensitive to the specific choice of the half-opening angle of the lanthanide-rich region $\Phi$ (for $15^\circ\lesssim\Phi\lesssim45^\circ$).

\section*{Conclusions and future perspectives}

In this work we have, for the first time, made quantitative predictions on the expected linear polarization signature from a kilonova produced by the coalescence of a binary NS system. We developed a simple model tailored to match the available observations of AT\,2017gfo\cite{Abbottetal2017b}, the only event of this category identified so far. The model does not include all the possible details that are currently under discussion in the literature (e.g. the bright ultraviolet emission detected for AT\,2017gfo a few hours after the merger\revised{\cite{Metzger2018}}) since they are either not expected to have an important effect on the derived polarization signature or their inclusion is still controversial\cite{Hotokezaka2013,Bauswein2013,Fernandez2013,Siegel2017,Shibata2017,Fujibayashi2017,Metzger2014,Lippuner2017}. Nevertheless, as far as polarization is concerned, our simple model proved to be flexible enough to allow us to effectively study the role of several parameters and derive fairly solid general conclusions. The model disregards the possible polarization contribution from the GRB afterglow since this is negligible during the kilonova evolution\cite{DAvanzoetal2018}. At the same time, however, the analysis of kilonova polarization is one of the best ways to look at the structure of NS-NS mergers from the angles at which we might not expect to see a GRB jet, but only the GW signal.

Kilonova polarization depends on the viewing angle of the source and is also time- and wavelength-dependent. The maximum value should be close to the $\sim 1$\% level, with some dependence on (as of yet) not fully constrained model parameters such as the angular distribution of the different ejecta and their expansion velocities. The polarization decreases going from equatorial to polar viewing angles $\theta_\mathrm{obs}$ thus providing an independent way to estimate this important parameter, which strongly impacts estimates of true rates from small numbers of detections (e.g. from future sources with less complete datasets than AT\,2017gfo). In the case of AT\,2017gfo, the viewing angle turns out to be smaller than $\sim 65^\circ$, \revised{a value that is less restrictive but consistent} with other independent estimates from the literature\cite{Abbottetal2017a,troja2017,Pianetal2017,Finstadetal2018,Mandel2018}. The polarization maximum should be detectable roughly in the R band due to the particular wavelength-dependence of bound-bound opacities combined to the presence of two ejecta components. No polarization is expected in the blue part of the optical spectrum \revised{as far as the ejecta are composed of r-process elements, regardless of whether or not they include lanthanides}. Already two or three days after the GW event, the intrinsic polarization from kilonovae should be virtually zero at any wavelength and viewing angle and this provides a reliable way to subtract the polarization possibly induced by dust along the line of sight. With just one observed event it is hard to evaluate how general the features derived from the observations can be. While a natural event-to-event variability is foreseeable, the general behaviour should anyway follow the results we have described here with some effect on the absolute polarization level. 

Polarimetry is one of the most valuable tools for uncovering the dynamics and physics of NS-NS mergers. For instance, detection of non-zero polarization in future events will unambiguously reveal the presence of lanthanide-free ejecta, unveil their spatial distribution -- which can not be investigated by any other means -- and thus allow us to study the merger dynamics responsible for synthesizing a diversity of r-process elements. At the same time, different results and likely higher polarization degrees are expected for a kilonova generated by the merging of a BH and a NS\revised{, where the lanthanide-rich dynamical ejecta are extremely deformed with an axial ratio of $\sim 0.2$} \revisedtwo{\cite{2013PhRvD..88d1503K,Kyutokuetal2015}}. This latter case requires a dedicated modelling and will be the target of a future study.

\begin{methods}

\section*{Expectations from hydrodynamical simulations of the binary merger}

We adopt models of kilonova ejecta which are motivated by both hydrodynamical simulations and the observed properties of the optical and near-infrared counterparts of GW170817, AT\,2017gfo. 

By hydrodynamical simulations, it is known that the ejecta from NS mergers have at least two components: (1) dynamical ejecta and (2) post-merger ejecta or disk wind. The first dynamical ejecta is driven by the tidal force as well as the shock heating\cite{Hotokezaka2013,Bauswein2013}. By the tidal disruption, the dynamical ejecta tend to be distributed around the equatorial plane. After the dynamical ejection, additional mass is ejected from the torus around the central object\cite{Fernandez2013,Siegel2017,Shibata2017,Fujibayashi2017}. The electron fraction $Y_e$ is an important parameter to determine the nucleosynthetic outcomes of the ejecta components. Because the shock heating tends to increase the value of $Y_e$, the dynamical ejecta can have broad distribution of $Y_e$. The post-merger ejecta can have higher $Y_e$, due to the irradiation from a neutron star at the center, although the exact $Y_e$ depends on the life time of a remnant neutron star\cite{Metzger2014,Lippuner2017}. In general, polar regions have higher $Y_e$. The mass, $Y_e$, and distribution of $Y_e$ in two components may vary according to the total mass and mass ratio of the merger.

If the ejecta include a substantial fraction of lanthanide elements, the kilonova emission would be faint and red due to the high opacities of lanthanides\cite{Kasenetal2013,Barnes&Kasen2013,Tanaka&Hotokezaka2013}. However, if the electron fraction is as high as $Y_e > 0.25$, production of lanthanide elements is suppressed\cite{Metzger2014,2015MNRAS.450.1777K,Tanakaetal2018}. \revised{Thus, the ejecta that consist only of such high $Y_e$ material produce brighter and bluer emission.} Therefore, the dynamical ejecta tend to give a faint, red kilonova while the post-merger ejecta are able to produce a bright, blue kilonova if the effects of neutrino irradiation is strong enough. It is noted that these connections may be too simple and an interplay between two components is also important\cite{2015MNRAS.450.1777K}, although in the present work we preferred to keep the scenario as simple as possible without introducing poorly constrained features.

\section*{Parameters for GW170817}

Multiple ejecta components have also been confirmed in AT 2017gfo. The observed light curves can be interpreted by the combination of the blue and red components\cite{Kasen2017,Tanakaetal2017}. The mass of each component is estimated to be $\sim 0.01-0.05$ M$_\odot$. Since the line of sight for GW170817 is $\lesssim30^\circ$ from the pole\cite{Abbottetal2017a,troja2017,Pianetal2017,Finstadetal2018,Mandel2018}, the blue component should exist in the polar direction without being fully absorbed by the lanthanide-rich ejecta. This is consistent with the expectations from simulations\revised{\cite{2017PhRvD..96l3012S}}. However, it is not yet clear if these blue and red components can be readily interpreted as the post-merger and dynamical ejecta, respectively. For example, the observed blue component has a velocity of $v \sim 0.2$~c\cite{Nicholl2017,Pianetal2017,Smarttetal2017}, which is higher than the expectation from post-merger ejecta\cite{Fernandez2013,Perego2014,Fujibayashi2017}. Also, the observed red component requires a mass\cite{Chornock2017,Cowperthwaite2017,Nicholl2017} larger than the typical mass from the dynamical mass ejection in the simulations\cite{Hotokezaka2013}.

\section*{Our model: densities, opacities and velocities}

Motivated by the simulations and the observations of AT 2017gfo, we adopt a simple model to calculate the polarization signals. \revised{The ejecta are assumed to be in homologous (free) expansion, i.e. material moving at velocity $v$ is located at a radius $r=v\,t$ at any given time $t$. This is a safe assumption for kilonovae already a few seconds after the merger.} The whole ejecta have $M_\mathrm{ej}=0.03\,M_\odot$ with a density structure of $r^{-3}$ from $v_\mathrm{in} = 0.05$~c to $v_\mathrm{out} = 0.3$~c. The slope of the density structure approximates the results of numerical simulations\revised{\cite{Tanaka&Hotokezaka2013,Hotokezaka2013}}. The photosphere is placed at $v_\mathrm{ph}=0.15$~c. The ejecta are assumed to be spherical. This is not necessarily true, in particular for the dynamical ejecta, but this choice does not have a large impact on the predicted polarization because a dominant part of the signal in our models results from the partial coverage of the photosphere (see Methods - Simulation results). The whole ejecta are divided into lanthanide-rich and lanthanide-free parts depending on the angle. The separation half-opening angle is taken to be $\Phi\sim30^\circ$ in our fiducial model. Since this angle is not well constrained by either theory or observations, we vary this parameter in our study and investigate its dependence on the predicted polarization (see Methods - Simulation results). 

\section*{Opacity calculations for lanthanide-rich and lanthanide-free ejecta}

We calculate wavelength-dependent opacities for our fiducial model using the time-dependent radiation transfer code \cite{Tanaka&Hotokezaka2013,Tanakaetal2018}. \revised{For lanthanide-rich ejecta as expected in the dynamical ejecta, we use the mass distribution of elements by evenly averaging the nucleosynthesis outcomes from $Y_e = 0.1$ to $Y_e = 0.4$. This approximates results of numerical simulations\cite{Tanakaetal2018}. Although the range includes a high $Y_e$, the final abundances include a large fraction of lanthanides ($\sim 9\%$ in mass fraction).} \revisedtwo{In the lanthanide-rich material, most of the opacity comes from singly or doubly ionized lanthanide elements with an open f-shell. In particular, the most important bound-bound transitions are those from low-lying ($\lesssim$~5~eV) energy levels.} \revised{For the lanthanide-free ejecta, we use element abundances from nucleosynthesis calculations with $Y_e = 0.3$ since the exact distribution of $Y_e$ in the disk ejecta has uncertainties.} Since the radial variation of the opacities is not strong above the photosphere, we simply use the opacities at the photosphere. On the other hand, the time evolution is quite strong reflecting a rapid change in the ionization states, and thus we use corresponding time snapshots for polarization calculations. Opacities at epochs corresponding to polarimetric observations of AT\,2017gfo (1.5, 2.5, 3.5 and 5.5~days) are shown in Fig.~\ref{fig:opac}. \revised{In particular, bound-bound opacities are averaged over a range of $\Delta\lambda=500$~\AA~centered at the desired wavelength (see Fig.~\ref{fig:opac_time_wave}).} Using this set of opacities, the polarization signals are calculated in a certain wavelength as a post process (see Methods - Polarization simulations).

\section*{Polarization simulations}

The three-dimensional Monte Carlo radiative transfer code \textsc{possis} (POlarization Spectral Synthesis In Supernovae) has been used in the past to model polarization spectra of Type Ia supernovae\cite{bulla2015,bulla2017} and superluminous supernovae\cite{inserra2016}. Here we extend its application to predict polarization signatures of kilonovae. \textsc{possis} adopts the description of the polarization state by the Stokes parameters $I$, $Q$, $U$, where $I$ is the total intensity and $Q$ and $U$ describe the linear polarization (the Stokes parameter describing circular polarization, $V$, is neglected in this work). Monte Carlo photon packets are created unpolarized ($Q=U=0$) on the spherical photosphere and emitted assuming constant surface brightness. Each photon packet streams freely throughout the ejecta until it undergoes either a Thomson (electron) scattering or a bound-bound line transition. Bound-free and free-free transitions are neglected as their opacities at epochs and wavelengths considered in this study are between $\sim$~5 and 12 orders of magnitudes smaller than electron scattering and bound-bound opacities. Thomson scattering is assumed to partially polarize the radiation, while bound-bound transitions are regarded as depolarizing contributions, \revised{a computational choice usually adopted in the literature}\cite{jeffery1989}. If an electron scattering event is selected\cite{Mazzali1993}, a new direction is sampled from the dipole Thomson scattering matrix and the Stokes parameters $I$, $Q$ and $U$ of the packet transformed accordingly\cite{chandrasekhar1960,Code1995}. If a line interaction is selected, the packet is instead re-emitted in a random direction with no polarization ($Q=U=0$). To compute polarization signals for different viewing angles, we use the ``event-based technique" (EBT) described in \cite{bulla2015} as this technique leads to a substantial reduction in the Monte Carlo noise compared to a direct-binning approach more usually adopted in Monte Carlo simulations. \revised{We note that rather smooth flux spectra as those observed in AT\,2017gfo\cite{Tanviretal2017} do not preclude rich polarization spectra\cite{1997ApJ...476L..27W}. However, opacity data used here (and in state-of-the-art simulations) are constructed from a few number atomic species, preventing us to predict reliable polarization spectra and draw any firm conclusion about polarization connected to individual line transitions.}

\section*{Simulation results}

In general, null polarization is predicted for spherical ejecta as each polarizing contribution is cancelled by another contribution one quadrant away\cite{Wang&Wheeler2008}. A net polarization signal can be reflecting either (i) an ejecta morphology departing from circular symmetry in projection or (ii) a non-homogeneous asymmetric opacity distribution within the whole ejecta. To explore (i), we compute polarization signatures \revised{at 7000~\AA}~for an ellipsoidal model with axial ratio 0.75 \revised{(in the range of theoretical expectations\cite{Hotokezaka2013})} and an homogeneous distribution of lanthanide-free opacities within the whole ejecta. When viewed from the equatorial plane, this system is associated with a polarization of $P_1=0.04\%$. Moving towards the pole, the ellipsoids become closer and closer to circular in projection and thus the predicted polarization levels becomes increasingly closer to zero. The predicted polarization levels are almost negligible due to the combination of small electron scattering opacities and of bound-bound line opacities of the same order ($\kappa_\mathrm{bb}\sim\kappa_\mathrm{es}\sim0.008$~cm$^{2}$~g$^{-1}$). For a comparison, polarization levels of $\sim2\%$ are predicted in supernova ellipsoidal models with same axial ratio, higher electron-scattering opacities and no line opacities\cite{1991A&A...246..481H}. \revised{Polarization levels similar to the one at 7000~\AA{} ($P\sim P_1\sim0.04\%$) are predicted at both 10\,000~and 15\,000~\AA, while increasingly smaller signals are found moving to shorter wavelengths ($P\sim0.005\%$ at 5000~\AA).} To explore (ii), we calculate polarization signatures \revised{at 7000~\AA} for a spherical model with lanthanide-rich opacities distributed around the equator (half-opening angle $\Phi=30^\circ$) and lanthanide-free opacities in the rest of the ejecta (see Fig.~\ref{fig:sketch}). Due to the asymmetric opacity distribution \revised{shown in Fig.~\ref{fig:opac}} ($\kappa_\mathrm{bb}^\mathrm{lf}\sim\,\kappa_\mathrm{es}^\mathrm{lf}\sim0.008$~cm$^{2}$~g$^{-1}$ in lanthanide-free while $\kappa_\mathrm{bb}^\mathrm{lr}\sim1500\,\kappa_\mathrm{es}^\mathrm{lr}\sim10$~cm$^{2}$~g$^{-1}$ in lanthanide-rich regions), the net polarization is no longer null as expected from spherical models but reaches values of $P_2=0.76\%$ 1.5~days after the merger for an equatorial orientation. Since $P_2>>P_1$, we conclude that the main source of polarization in kilonovae is given by the asymmetric distribution of lanthanide-rich material in the ejecta. For simplicity, the models presented in this study thus assume spherical ejecta.

Because radiation from the lanthanide-rich component is more likely to be depolarized by a line interaction before leaving the ejecta, most of the polarizing contributions to the signal come from regions at relatively high latitudes (i.e. closer to the poles). As shown in the inset of Fig.~\ref{fig:pol}, these contributions are preferentially associated with $Q<0$ values and this explains why the models presented in this work produces negative $Q$ polarization. Owing to the symmetry of the system about the equatorial plane, instead, $U=0$ in all the models ($U\neq0$ is expected for a different orientation of the equatorial-plane/merging-system on the sky). \revised{Deviations from zero in $U$ are associated with statistical fluctuations and used as a convenient proxy for the Monte Carlo noise in $Q$ (i.e. polarization values are quoted as $Q \pm |U|$).} As shown in Fig.~\ref{fig:pol}, the maximum polarization level ($Q\sim-0.8\%$) at 7000~\AA{} and at 1.5~days after the merger is reached for an observer in the equatorial plane ($\cos\theta_\mathrm{obs}=0$), while moving to higher latitudes leads to smaller absolute values since both the lanthanide-rich and lanthanide-free components become increasingly closer to circular symmetry in projection. As expected, $Q=0$ for an observer along the polar direction ($\cos\theta_\mathrm{obs}=1$).

The wavelength-dependence of the polarization signal at 1.5~days after the merger is shown in Fig.~\ref{fig:timewave}. In general, moving to both shorter and longer wavelengths than 7000~\AA{} results into smaller polarization levels. This is due to the combination of wavelength-dependent bound-bound opacities and the presence of two ejecta components. At shorter wavelengths (5000~\AA), bound-bound transitions become the dominant opacity source in both lanthanide-free ($\kappa_\mathrm{bb}^\mathrm{lf}\sim35\,\kappa_\mathrm{es}^\mathrm{lf}$) and lanthanide-rich ($\kappa_\mathrm{bb}^\mathrm{lr}\sim4200\,\kappa_\mathrm{es}^\mathrm{lr}$) components of the ejecta. Hence, depolarizing contributions dominate the signal and the predicted polarization level is rather small for all the viewing angles ($P\lesssim0.1\%$). \revised{Moving to longer wavelengths, instead, the relative importance of electron scattering to the total opacity increases in the lanthanide-rich ejecta ($\kappa_\mathrm{bb}^\mathrm{lr}\sim1500\,\kappa_\mathrm{es}$ at 7000~\AA, $\kappa_\mathrm{bb}^\mathrm{lr}\sim600\,\kappa_\mathrm{es}$ at 10\,000~\AA{} and $\kappa_\mathrm{bb}^\mathrm{lr}\sim150\,\kappa_\mathrm{es}$ at 15\,000~\AA). As a consequence, part of the polarizing contributions from the lanthanide-free region ($Q<0$) are cancelled by polarizing contributions one quadrant away in the lanthanide-rich region ($Q>0$), thus biasing the overall polarization signal towards smaller values than those at 7000~\AA.}


At 7000~\AA{} and 1.5~days after the merger, $\kappa_\mathrm{bb}^\mathrm{lf}\sim\kappa_\mathrm{es}^\mathrm{lf}$ in the lanthanide-free component while $\kappa_\mathrm{bb}^\mathrm{lr}\sim1500\,\kappa_\mathrm{es}^\mathrm{lr}$ in the lanthanide-rich component. Only one day after, bound-bound line transitions become the dominant source of opacities in all the ejecta, with $\kappa_\mathrm{bb}^\mathrm{lf}\sim4\,\kappa_\mathrm{es}^\mathrm{lf}$ and $\kappa_\mathrm{bb}^\mathrm{lr}\sim2800\,\kappa_\mathrm{es}^\mathrm{lr}$. Therefore, the rapid increase in bound-bound opacities -- and the small change in electron scattering opacities -- in both ejecta components \revised{(see Fig.~\ref{fig:opac_time_wave})} causes a very fast decrease in polarization between 1.5 and 2.5~d from the merger (i.e. the first two epochs of polarimetric observations of AT\,2017gfo). This is shown in Fig.~\ref{fig:timewave}, where the polarization along the equator at 7000~\AA{} changes from $P\sim0.8\%$ at 1.5~day to $P\sim0.1\%$ at 2.5~day. Later on, the polarization signal becomes negligible ($\kappa_\mathrm{bb}^\mathrm{lf}\sim400\,\kappa_\mathrm{es}^\mathrm{lf}$ and $\kappa_\mathrm{bb}^\mathrm{lr}\sim40\,000\,\kappa_\mathrm{es}^\mathrm{lr}$ at 3.5~days). Similar time-evolution are seen at different wavelengths.

Polarization in our models is attributed to the partial coverage of the photosphere by the high-opacity lanthanide-rich material. The extent of the lanthanide-rich component is therefore an important parameter in setting the absolute polarization values. In particular, different values of $\Phi$ correspond to different regions of the ejecta bringing low-polarization contributions and thus they lead to different polarization signals. Fig.~\ref{fig:opang} shows polarization levels at 1.5~days for different values of $\Phi$: 15$^\circ$, 22.5$^\circ$, 30$^\circ$, 37.5$^\circ$ and 45$^\circ$.  The maximum polarization level is reached for $\Phi= 30^\circ$ since in this configuration the partial coverage of the photosphere by the lanthanide-rich ejecta corresponds to half of the full solid angle, i.e. $\Delta\Omega=4\pi\times\sin\Phi=2\pi$ (see inset of Fig.~\ref{fig:pol}). For lower values of $\Phi$, the partial coverage of the photosphere is smaller and positive $Q$ contributions from regions close to the equatorial plane partially cancel the negative contribution from polar regions, thus bringing the overall polarization signal to smaller absolute values. For higher values of $\Phi$, instead, the partial coverage is higher, causing an increase in the contribution to the total flux of photons from the lanthanide-rich regions. Since these photon packets are preferentially unpolarized, the overall polarization level decreases. Nevertheless, polarization signals for $\cos\theta_\mathrm{obs}\gtrsim0.4$ are predicted to be similar for all the set of $\Phi$ values investigated since at relatively high latitudes the lanthanide-rich component is deeper inside the ejecta \revised{(see Fig.~\ref{fig:sketch})} and contributes less to the final emitted flux. \revised{Although hydrodynamical simulations\cite{2015MNRAS.450.1777K,2015ApJ...813....2M,2017PhRvD..96l3012S,2017PhRvD..96l4005B} suggest that $\Phi$ might be in the range of 30$-$45~degrees, we note that larger values -- not excluded \textit{per se} -- would be associated to negligible polarization signals ($P\lesssim0.2\%$).}

The location of the photosphere (parametrised by the photospheric velocity $v_\mathrm{ph}$) has a relatively large impact on the polarization signal. In our fiducial model, with the photosphere modelled as a spherical surface at $v_\mathrm{ph}=0.15$~c, the polarization in the equatorial plane is $P\sim0.8\%$ at 7000~\AA{} and 1.5~days after the merger. Moving the photosphere to $v_\mathrm{ph}=0.1$~c, the relative increase in optical depth to the boundary is larger in the lanthanide-rich compared to the lanthanide-free regions. This means that the contribution of the lanthanide-rich region to the total flux changes from $\sim$~93~per~cent at $v_\mathrm{ph}=0.15$~c to $\sim$~89~per~cent at $v_\mathrm{ph}=0.1$~c. Because of the depolarizing contribution of the lanthanide-rich component, the decrease in flux corresponds to an increase in the overall polarization. Specifically, a polarization of $P\sim1.9\%$ is found for the model with $v_\mathrm{ph}=0.1$~c, roughly twice as large compared to the fiducial model with $v_\mathrm{ph}=0.15$~c.


\section*{Comparison with previous polarimetric studies}

\revisedtwo{Polarization signatures for similar geometries have been predicted in previous stellar\cite{1994A&A...289..492H,1996ApJ...461..847W} and supernova\cite{2003ApJ...593..788K,2011MNRAS.415.3497D,bulla2015} studies. The qualitative polarization behaviours identified in these works bear strong similarities to those found in our simulations. First, a polarization decrease to zero with increasing inclination (see Fig.~\ref{fig:pol}) is predicted by previous studies adopting similar axisymmetric morphologies (see e.g. fig.~5b of \cite{1996ApJ...461..847W}, fig.~18 of \cite{2011MNRAS.415.3497D} and fig.~7 of \cite{bulla2015}). Moreover, the temporal evolution of polarization seen in Fig.~\ref{fig:timewave} is in qualitative agreement with that predicted by axisymmetric supernova models although over different time-scales (see e.g. fig. 16 of \cite{bulla2015}). Finally, the smaller polarization levels found at 5000 compared to 7000~\AA{} (see Fig.~\ref{fig:timewave}) are reminiscent of those predicted in Type Ia supernova models and usually ascribed to the increasing importance of depolarizing bound-bound transitions when moving from longer to shorter optical wavelengths\cite{2004ApJ...610..876K,2016MNRAS.462.1039B}.}

\revisedtwo{Differences in the adopted morphologies, opacities and densities make quantitative comparisons between polarization levels from our simulations and those from the literature difficult. For instance, studies like \cite{1994A&A...289..492H,1996ApJ...461..847W} neglect bound-bound opacities and focus on continuum processes by predicting the polarization signals as a function of the albedo $a=\kappa_\mathrm{es}/\kappa_\mathrm{tot}=\kappa_\mathrm{es}/(\kappa_\mathrm{es}+\kappa_\mathrm{bf}+\kappa_\mathrm{ff})$, where $\kappa_\mathrm{bf}$ and $\kappa_\mathrm{ff}$ are the bound-free and free-free opacities, respectively. In contrast, bound-free and free-free opacities are neglected in our simulations and the albedo computed as $\kappa_\mathrm{es}/(\kappa_\mathrm{es}+\kappa_\mathrm{bb})$. More importantly, previous studies remove photons undergoing either bound-free or free-free interactions from the simulations. In contrast, no sink is introduced in our calculations and photons undergoing bound-bound interactions are re-emitted isotropically and with no polarization. The different choices in the opacity sources and treatments make a significant impact on the predicted polarization levels.}

\revisedtwo{Although a full quantitative comparison is beyond the scope of this study, we note that polarization percentages predicted here are in the same range found by previous works. For instance, our simulations adopt albedo of $\sim$~0 and $\sim$~0.5 at 1.5~days and at 7000~\AA{} in the lanthanide-rich and lanthanide-free ejecta, respectively (see Methods - Simulation results). The resulting polarization level of $\sim$~0.8\,\% for an equatorial viewing angle (see Fig.~\ref{fig:pol}) is within the range of $\sim$~$0.2-1.5$\,\% predicted by \cite{1996ApJ...461..847W} for albedo between 0.1 and 0.5 (see their fig. 5b).}

\section*{Upper limits to AT\,2017gfo intrinsic polarization}
The results of the polarimetric campaign carried out for AT\,2017gfo are reported in \cite{Covinoetal2017}. In spite of the fair signal-to-noise obtained with the observations, the main limitation to the derived upper limits, $P<0.4-0.5$\,\%, is due to the effect of polarization induced by dust along the line of sight in both our and the host galaxy\cite{Serkowskietal1975}. In fact, observation of stars in the field of view showed polarization from virtually zero up to $\sim 0.7$\,\%. The polarization derived for AT\,2017gfo during the first epoch ($\sim 1.5$\,days after the event) is consistent with this value, while at later epoch only upper limits were obtained. Following a strategy already applied for supernova spectro-polarimetric observations (e.g. \cite{Maund2007,Patatetal2012}), it is possible to derive a much better estimate for the first epoch polarization. According to the results of our analysis (see Polarization), the R band flux from the kilonova is completely non polarized for any viewing direction at epochs later than $2-3$~days, due to the high line opacity of the emitting material. Therefore, by a weighted average of the results of the observations in the R band at all epochs but the first we can obtain a solid estimate of the polarization induced by any dust along the line of sight. Using data reported in Tables\,1 and 2 in \cite{Covinoetal2017} we get $Q_{\rm dust} = -0.0021 \pm 0.0006$ and $U_{\rm dust} = 0.0044 \pm 0.0005$, corresponding to $P_{\rm dust} = (0.49 \pm 0.05)$\% and a position angle $\chi_{\rm dust} = (58 \pm 4)^\circ$, in fair agreement with what it was observed for the field stars\cite{Covinoetal2017}. By subtracting the average value to $Q$ and $U$ obtained for the first epoch we then derive our best estimate of the intrinsic polarization from AT\,2017gfo: $Q = -0.0003 \pm 0.0009$ and $U = 0.0009 \pm 0.0005$, which corresponds, following \cite{Plaszczynskiet2014}, to an upper limit of $P < 0.18$\% at 95\% confidence level.

\paragraph{Code availability.} The radiative transfer code \textsc{possis} used in this work is not publicly available. Results presented in this work are available from the corresponding author upon reasonable request.

\end{methods}



\begin{thebibliography}{}
\expandafter\ifx\csname url\endcsname\relax
  \def\url#1{\texttt{#1}}\fi
\expandafter\ifx\csname urlprefix\endcsname\relax\def\urlprefix{URL }\fi
\providecommand{\bibinfo}[2]{#2}
\providecommand{\eprint}[2][]{\url{#2}}

\end{thebibliography}


\begin{thebibliography}{10}
\expandafter\ifx\csname url\endcsname\relax
  \def\url#1{\texttt{#1}}\fi
\expandafter\ifx\csname urlprefix\endcsname\relax\def\urlprefix{URL }\fi
\providecommand{\bibinfo}[2]{#2}
\providecommand{\eprint}[2][]{\url{#2}}

\bibitem{Abbottetal2017a}
\bibinfo{author}{{Abbott}, B.~P.} \emph{et~al.}
\newblock \bibinfo{title}{{GW170817: Observation of Gravitational Waves from a
  Binary Neutron Star Inspiral}}.
\newblock \emph{\bibinfo{journal}{Physical Review Letters}}
  \textbf{\bibinfo{volume}{119}}, \bibinfo{pages}{161101}
  (\bibinfo{year}{2017}).
\newblock \eprint{1710.05832}.

\bibitem{Abbottetal2017b}
\bibinfo{author}{{Abbott}, B.~P.} \emph{et~al.}
\newblock \bibinfo{title}{{Multi-messenger Observations of a Binary Neutron
  Star Merger}}.
\newblock \emph{\bibinfo{journal}{\apjl}} \textbf{\bibinfo{volume}{848}},
  \bibinfo{pages}{L12} (\bibinfo{year}{2017}).
\newblock \eprint{1710.05833}.

\bibitem{Mooleyetal2018b}
\bibinfo{author}{{Mooley}, K.~P.} \emph{et~al.}
\newblock \bibinfo{title}{{Superluminal motion of a relativistic jet in the
  neutron star merger GW170817}}.
\newblock \emph{\bibinfo{journal}{ArXiv e-prints}}  (\bibinfo{year}{2018}).
\newblock \eprint{1806.09693}.

\bibitem{Ghirlandaetal2018}
\bibinfo{author}{{Ghirlanda}, G.} \emph{et~al.}
\newblock \bibinfo{title}{{Re-solving the jet/cocoon riddle of the first
  gravitational wave with an electromagnetic counterpart}}.
\newblock \emph{\bibinfo{journal}{ArXiv e-prints}}  (\bibinfo{year}{2018}).
\newblock \eprint{1808.00469}.

\bibitem{Covinoetal2017}
\bibinfo{author}{{Covino}, S.} \emph{et~al.}
\newblock \bibinfo{title}{{The unpolarized macronova associated with the
  gravitational wave event GW 170817}}.
\newblock \emph{\bibinfo{journal}{Nature Astronomy}}
  \textbf{\bibinfo{volume}{1}}, \bibinfo{pages}{791--794}
  (\bibinfo{year}{2017}).
\newblock \eprint{1710.05849}.

\bibitem{Evansetal2017}
\bibinfo{author}{{Evans}, P.~A.} \emph{et~al.}
\newblock \bibinfo{title}{{Swift and NuSTAR observations of GW170817: Detection
  of a blue kilonova}}.
\newblock \emph{\bibinfo{journal}{Science}} \textbf{\bibinfo{volume}{358}},
  \bibinfo{pages}{1565--1570} (\bibinfo{year}{2017}).
\newblock \eprint{1710.05437}.

\bibitem{Pianetal2017}
\bibinfo{author}{{Pian}, E.} \emph{et~al.}
\newblock \bibinfo{title}{{Spectroscopic identification of r-process
  nucleosynthesis in a double neutron-star merger}}.
\newblock \emph{\bibinfo{journal}{\nat}} \textbf{\bibinfo{volume}{551}},
  \bibinfo{pages}{67--70} (\bibinfo{year}{2017}).
\newblock \eprint{1710.05858}.

\bibitem{Smarttetal2017}
\bibinfo{author}{{Smartt}, S.~J.} \emph{et~al.}
\newblock \bibinfo{title}{{A kilonova as the electromagnetic counterpart to a
  gravitational-wave source}}.
\newblock \emph{\bibinfo{journal}{\nat}} \textbf{\bibinfo{volume}{551}},
  \bibinfo{pages}{75--79} (\bibinfo{year}{2017}).
\newblock \eprint{1710.05841}.

\bibitem{Tanviretal2017}
\bibinfo{author}{{Tanvir}, N.~R.} \emph{et~al.}
\newblock \bibinfo{title}{{The Emergence of a Lanthanide-rich Kilonova
  Following the Merger of Two Neutron Stars}}.
\newblock \emph{\bibinfo{journal}{\apjl}} \textbf{\bibinfo{volume}{848}},
  \bibinfo{pages}{L27} (\bibinfo{year}{2017}).
\newblock \eprint{1710.05455}.

\bibitem{Li&Pacynski1998}
\bibinfo{author}{{Li}, L.-X.} \& \bibinfo{author}{{Paczy{\'n}ski}, B.}
\newblock \bibinfo{title}{{Transient Events from Neutron Star Mergers}}.
\newblock \emph{\bibinfo{journal}{\apjl}} \textbf{\bibinfo{volume}{507}},
  \bibinfo{pages}{L59--L62} (\bibinfo{year}{1998}).
\newblock \eprint{astro-ph/9807272}.

\bibitem{Metzgeretal2010}
\bibinfo{author}{{Metzger}, B.~D.} \emph{et~al.}
\newblock \bibinfo{title}{{Electromagnetic counterparts of compact object
  mergers powered by the radioactive decay of r-process nuclei}}.
\newblock \emph{\bibinfo{journal}{\mnras}} \textbf{\bibinfo{volume}{406}},
  \bibinfo{pages}{2650--2662} (\bibinfo{year}{2010}).
\newblock \eprint{1001.5029}.

\bibitem{Roberts2011}
\bibinfo{author}{{Roberts}, L.~F.}, \bibinfo{author}{{Kasen}, D.},
  \bibinfo{author}{{Lee}, W.~H.} \& \bibinfo{author}{{Ramirez-Ruiz}, E.}
\newblock \bibinfo{title}{{Electromagnetic Transients Powered by Nuclear Decay
  in the Tidal Tails of Coalescing Compact Binaries}}.
\newblock \emph{\bibinfo{journal}{\apjl}} \textbf{\bibinfo{volume}{736}},
  \bibinfo{pages}{L21} (\bibinfo{year}{2011}).
\newblock \eprint{1104.5504}.

\bibitem{Kasenetal2013}
\bibinfo{author}{{Kasen}, D.}, \bibinfo{author}{{Badnell}, N.~R.} \&
  \bibinfo{author}{{Barnes}, J.}
\newblock \bibinfo{title}{{Opacities and Spectra of the r-process Ejecta from
  Neutron Star Mergers}}.
\newblock \emph{\bibinfo{journal}{\apj}} \textbf{\bibinfo{volume}{774}},
  \bibinfo{pages}{25} (\bibinfo{year}{2013}).
\newblock \eprint{1303.5788}.

\bibitem{Barnes&Kasen2013}
\bibinfo{author}{{Barnes}, J.} \& \bibinfo{author}{{Kasen}, D.}
\newblock \bibinfo{title}{{Effect of a High Opacity on the Light Curves of
  Radioactively Powered Transients from Compact Object Mergers}}.
\newblock \emph{\bibinfo{journal}{\apj}} \textbf{\bibinfo{volume}{775}},
  \bibinfo{pages}{18} (\bibinfo{year}{2013}).
\newblock \eprint{1303.5787}.

\bibitem{Tanaka&Hotokezaka2013}
\bibinfo{author}{{Tanaka}, M.} \& \bibinfo{author}{{Hotokezaka}, K.}
\newblock \bibinfo{title}{{Radiative Transfer Simulations of Neutron Star
  Merger Ejecta}}.
\newblock \emph{\bibinfo{journal}{\apj}} \textbf{\bibinfo{volume}{775}},
  \bibinfo{pages}{113} (\bibinfo{year}{2013}).
\newblock \eprint{1306.3742}.

\bibitem{Baiotti&Rezzolla2017}
\bibinfo{author}{{Baiotti}, L.} \& \bibinfo{author}{{Rezzolla}, L.}
\newblock \bibinfo{title}{{Binary neutron star mergers: a review of Einstein's
  richest laboratory}}.
\newblock \emph{\bibinfo{journal}{Reports on Progress in Physics}}
  \textbf{\bibinfo{volume}{80}}, \bibinfo{pages}{096901}
  (\bibinfo{year}{2017}).
\newblock \eprint{1607.03540}.

\bibitem{Metzger2017}
\bibinfo{author}{{Metzger}, B.~D.}
\newblock \bibinfo{title}{{Kilonovae}}.
\newblock \emph{\bibinfo{journal}{Living Reviews in Relativity}}
  \textbf{\bibinfo{volume}{20}}, \bibinfo{pages}{3} (\bibinfo{year}{2017}).
\newblock \eprint{1610.09381}.

\bibitem{Tanakaetal2017}
\bibinfo{author}{{Tanaka}, M.} \emph{et~al.}
\newblock \bibinfo{title}{{Kilonova from post-merger ejecta as an optical and
  near-Infrared counterpart of GW170817}}.
\newblock \emph{\bibinfo{journal}{\pasj}} \textbf{\bibinfo{volume}{69}},
  \bibinfo{pages}{102} (\bibinfo{year}{2017}).
\newblock \eprint{1710.05850}.

\bibitem{Tanakaetal2018}
\bibinfo{author}{{Tanaka}, M.} \emph{et~al.}
\newblock \bibinfo{title}{{Properties of Kilonovae from Dynamical and
  Post-merger Ejecta of Neutron Star Mergers}}.
\newblock \emph{\bibinfo{journal}{\apj}} \textbf{\bibinfo{volume}{852}},
  \bibinfo{pages}{109} (\bibinfo{year}{2018}).
\newblock \eprint{1708.09101}.

\bibitem{2013PhRvD..88d1503K}
\bibinfo{author}{{Kyutoku}, K.}, \bibinfo{author}{{Ioka}, K.} \&
  \bibinfo{author}{{Shibata}, M.}
\newblock \bibinfo{title}{{Anisotropic mass ejection from black hole-neutron
  star binaries: Diversity of electromagnetic counterparts}}.
\newblock \emph{\bibinfo{journal}{\prd}} \textbf{\bibinfo{volume}{88}},
  \bibinfo{pages}{041503} (\bibinfo{year}{2013}).
\newblock \eprint{1305.6309}.

\bibitem{Kyutokuetal2015}
\bibinfo{author}{{Kyutoku}, K.}, \bibinfo{author}{{Ioka}, K.},
  \bibinfo{author}{{Okawa}, H.}, \bibinfo{author}{{Shibata}, M.} \&
  \bibinfo{author}{{Taniguchi}, K.}
\newblock \bibinfo{title}{{Dynamical mass ejection from black hole-neutron star
  binaries}}.
\newblock \emph{\bibinfo{journal}{\prd}} \textbf{\bibinfo{volume}{92}},
  \bibinfo{pages}{044028} (\bibinfo{year}{2015}).
\newblock \eprint{1502.05402}.

\bibitem{Hotokezaka2013}
\bibinfo{author}{{Hotokezaka}, K.} \emph{et~al.}
\newblock \bibinfo{title}{{Mass ejection from the merger of binary neutron
  stars}}.
\newblock \emph{\bibinfo{journal}{\prd}} \textbf{\bibinfo{volume}{87}},
  \bibinfo{pages}{024001} (\bibinfo{year}{2013}).
\newblock \eprint{1212.0905}.

\bibitem{Bauswein2013}
\bibinfo{author}{{Bauswein}, A.}, \bibinfo{author}{{Goriely}, S.} \&
  \bibinfo{author}{{Janka}, H.-T.}
\newblock \bibinfo{title}{{Systematics of Dynamical Mass Ejection,
  Nucleosynthesis, and Radioactively Powered Electromagnetic Signals from
  Neutron-star Mergers}}.
\newblock \emph{\bibinfo{journal}{\apj}} \textbf{\bibinfo{volume}{773}},
  \bibinfo{pages}{78} (\bibinfo{year}{2013}).
\newblock \eprint{1302.6530}.

\bibitem{Fernandez2013}
\bibinfo{author}{{Fern{\'a}ndez}, R.} \& \bibinfo{author}{{Metzger}, B.~D.}
\newblock \bibinfo{title}{{Delayed outflows from black hole accretion tori
  following neutron star binary coalescence}}.
\newblock \emph{\bibinfo{journal}{\mnras}} \textbf{\bibinfo{volume}{435}},
  \bibinfo{pages}{502--517} (\bibinfo{year}{2013}).
\newblock \eprint{1304.6720}.

\bibitem{Siegel2017}
\bibinfo{author}{{Siegel}, D.~M.} \& \bibinfo{author}{{Metzger}, B.~D.}
\newblock \bibinfo{title}{{Three-Dimensional General-Relativistic
  Magnetohydrodynamic Simulations of Remnant Accretion Disks from Neutron Star
  Mergers: Outflows and r -Process Nucleosynthesis}}.
\newblock \emph{\bibinfo{journal}{Physical Review Letters}}
  \textbf{\bibinfo{volume}{119}}, \bibinfo{pages}{231102}
  (\bibinfo{year}{2017}).
\newblock \eprint{1705.05473}.

\bibitem{Shibata2017}
\bibinfo{author}{{Shibata}, M.}, \bibinfo{author}{{Kiuchi}, K.} \&
  \bibinfo{author}{{Sekiguchi}, Y.-i.}
\newblock \bibinfo{title}{{General relativistic viscous hydrodynamics of
  differentially rotating neutron stars}}.
\newblock \emph{\bibinfo{journal}{\prd}} \textbf{\bibinfo{volume}{95}},
  \bibinfo{pages}{083005} (\bibinfo{year}{2017}).
\newblock \eprint{1703.10303}.

\bibitem{Fujibayashi2017}
\bibinfo{author}{{Fujibayashi}, S.}, \bibinfo{author}{{Sekiguchi}, Y.},
  \bibinfo{author}{{Kiuchi}, K.} \& \bibinfo{author}{{Shibata}, M.}
\newblock \bibinfo{title}{{Properties of Neutrino-driven Ejecta from the
  Remnant of Binary Neutron Star Merger : Purely Radiation Hydrodynamics
  Case}}.
\newblock \emph{\bibinfo{journal}{ArXiv e-prints}}  (\bibinfo{year}{2017}).
\newblock \eprint{1703.10191}.

\bibitem{Metzger2014}
\bibinfo{author}{{Metzger}, B.~D.} \& \bibinfo{author}{{Fern{\'a}ndez}, R.}
\newblock \bibinfo{title}{{Red or blue? A potential kilonova imprint of the
  delay until black hole formation following a neutron star merger}}.
\newblock \emph{\bibinfo{journal}{\mnras}} \textbf{\bibinfo{volume}{441}},
  \bibinfo{pages}{3444--3453} (\bibinfo{year}{2014}).
\newblock \eprint{1402.4803}.

\bibitem{Lippuner2017}
\bibinfo{author}{{Lippuner}, J.} \emph{et~al.}
\newblock \bibinfo{title}{{Signatures of hypermassive neutron star lifetimes on
  r-process nucleosynthesis in the disc ejecta from neutron star mergers}}.
\newblock \emph{\bibinfo{journal}{\mnras}} \textbf{\bibinfo{volume}{472}},
  \bibinfo{pages}{904--918} (\bibinfo{year}{2017}).
\newblock \eprint{1703.06216}.

\bibitem{Perego2014}
\bibinfo{author}{{Perego}, A.} \emph{et~al.}
\newblock \bibinfo{title}{{Neutrino-driven winds from neutron star merger
  remnants}}.
\newblock \emph{\bibinfo{journal}{\mnras}} \textbf{\bibinfo{volume}{443}},
  \bibinfo{pages}{3134--3156} (\bibinfo{year}{2014}).
\newblock \eprint{1405.6730}.

\bibitem{Chornock2017}
\bibinfo{author}{{Chornock}, R.} \emph{et~al.}
\newblock \bibinfo{title}{{The Electromagnetic Counterpart of the Binary
  Neutron Star Merger LIGO/Virgo GW170817. IV. Detection of Near-infrared
  Signatures of r-process Nucleosynthesis with Gemini-South}}.
\newblock \emph{\bibinfo{journal}{\apjl}} \textbf{\bibinfo{volume}{848}},
  \bibinfo{pages}{L19} (\bibinfo{year}{2017}).
\newblock \eprint{1710.05454}.

\bibitem{Cowperthwaite2017}
\bibinfo{author}{{Cowperthwaite}, P.~S.} \emph{et~al.}
\newblock \bibinfo{title}{{The Electromagnetic Counterpart of the Binary
  Neutron Star Merger LIGO/Virgo GW170817. II. UV, Optical, and Near-infrared
  Light Curves and Comparison to Kilonova Models}}.
\newblock \emph{\bibinfo{journal}{\apjl}} \textbf{\bibinfo{volume}{848}},
  \bibinfo{pages}{L17} (\bibinfo{year}{2017}).
\newblock \eprint{1710.05840}.

\bibitem{Nicholl2017}
\bibinfo{author}{{Nicholl}, M.} \emph{et~al.}
\newblock \bibinfo{title}{{The Electromagnetic Counterpart of the Binary
  Neutron Star Merger LIGO/Virgo GW170817. III. Optical and UV Spectra of a
  Blue Kilonova from Fast Polar Ejecta}}.
\newblock \emph{\bibinfo{journal}{\apjl}} \textbf{\bibinfo{volume}{848}},
  \bibinfo{pages}{L18} (\bibinfo{year}{2017}).
\newblock \eprint{1710.05456}.

\bibitem{bulla2015}
\bibinfo{author}{{Bulla}, M.}, \bibinfo{author}{{Sim}, S.~A.} \&
  \bibinfo{author}{{Kromer}, M.}
\newblock \bibinfo{title}{{Polarization spectral synthesis for Type Ia
  supernova explosion models}}.
\newblock \emph{\bibinfo{journal}{\mnras}} \textbf{\bibinfo{volume}{450}},
  \bibinfo{pages}{967--981} (\bibinfo{year}{2015}).
\newblock \eprint{1503.07002}.

\bibitem{bulla2017}
\bibinfo{author}{{Bulla}, M.}
\newblock \emph{\bibinfo{title}{{Polarisation Spectral Synthesis For Type Ia
  Supernova Explosion Models}}}.
\newblock Ph.D. thesis, \bibinfo{school}{Astrophysics Research Centre, School
  of Mathematics and Physics, Queen's University Belfast, Belfast BT7 1NN, UK}
  (\bibinfo{year}{2017}).

\bibitem{Serkowskietal1975}
\bibinfo{author}{{Serkowski}, K.}, \bibinfo{author}{{Mathewson}, D.~S.} \&
  \bibinfo{author}{{Ford}, V.~L.}
\newblock \bibinfo{title}{{Wavelength dependence of interstellar polarization
  and ratio of total to selective extinction}}.
\newblock \emph{\bibinfo{journal}{\apj}} \textbf{\bibinfo{volume}{196}},
  \bibinfo{pages}{261--290} (\bibinfo{year}{1975}).

\bibitem{Wang&Wheeler2008}
\bibinfo{author}{{Wang}, L.} \& \bibinfo{author}{{Wheeler}, J.~C.}
\newblock \bibinfo{title}{{Spectropolarimetry of Supernovae}}.
\newblock \emph{\bibinfo{journal}{\araa}} \textbf{\bibinfo{volume}{46}},
  \bibinfo{pages}{433--474} (\bibinfo{year}{2008}).
\newblock \eprint{0811.1054}.

\bibitem{Maund2007}
\bibinfo{author}{{Maund}, J.~R.} \emph{et~al.}
\newblock \bibinfo{title}{{Spectropolarimetry of the Type IIb Supernova
  2001ig}}.
\newblock \emph{\bibinfo{journal}{\apj}} \textbf{\bibinfo{volume}{671}},
  \bibinfo{pages}{1944--1958} (\bibinfo{year}{2007}).
\newblock \eprint{0709.1487}.

\bibitem{Patatetal2012}
\bibinfo{author}{{Patat}, F.} \emph{et~al.}
\newblock \bibinfo{title}{{VLT Spectropolarimetry of the Type Ia SN 2005ke. A
  step towards understanding subluminous events}}.
\newblock \emph{\bibinfo{journal}{\aap}} \textbf{\bibinfo{volume}{545}},
  \bibinfo{pages}{A7} (\bibinfo{year}{2012}).
\newblock \eprint{1206.1858}.

\bibitem{troja2017}
\bibinfo{author}{{Troja}, E.} \emph{et~al.}
\newblock \bibinfo{title}{{The X-ray counterpart to the gravitational-wave
  event GW170817}}.
\newblock \emph{\bibinfo{journal}{\nat}} \textbf{\bibinfo{volume}{551}},
  \bibinfo{pages}{71--74} (\bibinfo{year}{2017}).
\newblock \eprint{1710.05433}.

\bibitem{Finstadetal2018}
\bibinfo{author}{{Finstad}, D.}, \bibinfo{author}{{De}, S.},
  \bibinfo{author}{{Brown}, D.~A.}, \bibinfo{author}{{Berger}, E.} \&
  \bibinfo{author}{{Biwer}, C.~M.}
\newblock \bibinfo{title}{{Measuring the viewing angle of GW170817 with
  electromagnetic and gravitational waves}}.
\newblock \emph{\bibinfo{journal}{ArXiv e-prints}}  (\bibinfo{year}{2018}).
\newblock \eprint{1804.04179}.

\bibitem{Mandel2018}
\bibinfo{author}{{Mandel}, I.}
\newblock \bibinfo{title}{{The Orbit of GW170817 Was Inclined by Less Than
  $28{\deg}$ to the Line of Sight}}.
\newblock \emph{\bibinfo{journal}{\apjl}} \textbf{\bibinfo{volume}{853}},
  \bibinfo{pages}{L12} (\bibinfo{year}{2018}).
\newblock \eprint{1712.03958}.

\bibitem{Metzger2018}
\bibinfo{author}{{Metzger}, B.~D.}, \bibinfo{author}{{Thompson}, T.~A.} \&
  \bibinfo{author}{{Quataert}, E.}
\newblock \bibinfo{title}{{A Magnetar Origin for the Kilonova Ejecta in
  GW170817}}.
\newblock \emph{\bibinfo{journal}{\apj}} \textbf{\bibinfo{volume}{856}},
  \bibinfo{pages}{101} (\bibinfo{year}{2018}).
\newblock \eprint{1801.04286}.

\bibitem{DAvanzoetal2018}
\bibinfo{author}{{D'Avanzo}, P.} \emph{et~al.}
\newblock \bibinfo{title}{{The evolution of the X-ray afterglow emission of GW
  170817 / GRB 170817A in XMM-Newton observations}}.
\newblock \emph{\bibinfo{journal}{ArXiv e-prints}}  (\bibinfo{year}{2018}).
\newblock \eprint{1801.06164}.

\bibitem{2015MNRAS.450.1777K}
\bibinfo{author}{{Kasen}, D.}, \bibinfo{author}{{Fern{\'a}ndez}, R.} \&
  \bibinfo{author}{{Metzger}, B.~D.}
\newblock \bibinfo{title}{{Kilonova light curves from the disc wind outflows of
  compact object mergers}}.
\newblock \emph{\bibinfo{journal}{\mnras}} \textbf{\bibinfo{volume}{450}},
  \bibinfo{pages}{1777--1786} (\bibinfo{year}{2015}).
\newblock \eprint{1411.3726}.

\bibitem{Kasen2017}
\bibinfo{author}{{Kasen}, D.}, \bibinfo{author}{{Metzger}, B.},
  \bibinfo{author}{{Barnes}, J.}, \bibinfo{author}{{Quataert}, E.} \&
  \bibinfo{author}{{Ramirez-Ruiz}, E.}
\newblock \bibinfo{title}{{Origin of the heavy elements in binary neutron-star
  mergers from a gravitational-wave event}}.
\newblock \emph{\bibinfo{journal}{\nat}} \textbf{\bibinfo{volume}{551}},
  \bibinfo{pages}{80--84} (\bibinfo{year}{2017}).
\newblock \eprint{1710.05463}.

\bibitem{2017PhRvD..96l3012S}
\bibinfo{author}{{Shibata}, M.} \emph{et~al.}
\newblock \bibinfo{title}{{Modeling GW170817 based on numerical relativity and
  its implications}}.
\newblock \emph{\bibinfo{journal}{\prd}} \textbf{\bibinfo{volume}{96}},
  \bibinfo{pages}{123012} (\bibinfo{year}{2017}).
\newblock \eprint{1710.07579}.

\bibitem{inserra2016}
\bibinfo{author}{{Inserra}, C.}, \bibinfo{author}{{Bulla}, M.},
  \bibinfo{author}{{Sim}, S.~A.} \& \bibinfo{author}{{Smartt}, S.~J.}
\newblock \bibinfo{title}{{Spectropolarimetry of Superluminous Supernovae:
  Insight into Their Geometry}}.
\newblock \emph{\bibinfo{journal}{\apj}} \textbf{\bibinfo{volume}{831}},
  \bibinfo{pages}{79} (\bibinfo{year}{2016}).
\newblock \eprint{1607.02353}.

\bibitem{jeffery1989}
\bibinfo{author}{{Jeffery}, D.~J.}
\newblock \bibinfo{title}{{The Sobolev-P method - A generalization of the
  Sobolev method for the treatment of the polarization state of radiation and
  the polarizing effect of resonance line scattering}}.
\newblock \emph{\bibinfo{journal}{\apjs}} \textbf{\bibinfo{volume}{71}},
  \bibinfo{pages}{951--981} (\bibinfo{year}{1989}).

\bibitem{Mazzali1993}
\bibinfo{author}{{Mazzali}, P.~A.} \& \bibinfo{author}{{Lucy}, L.~B.}
\newblock \bibinfo{title}{{The application of Monte Carlo methods to the
  synthesis of early-time supernovae spectra}}.
\newblock \emph{\bibinfo{journal}{\aap}} \textbf{\bibinfo{volume}{279}},
  \bibinfo{pages}{447--456} (\bibinfo{year}{1993}).

\bibitem{chandrasekhar1960}
\bibinfo{author}{{Chandrasekhar}, S.}
\newblock \emph{\bibinfo{title}{{Radiative transfer}}}
  (\bibinfo{publisher}{Dover Publications, New York}, \bibinfo{year}{1960}).

\bibitem{Code1995}
\bibinfo{author}{{Code}, A.~D.} \& \bibinfo{author}{{Whitney}, B.~A.}
\newblock \bibinfo{title}{{Polarization from scattering in blobs}}.
\newblock \emph{\bibinfo{journal}{\apj}} \textbf{\bibinfo{volume}{441}},
  \bibinfo{pages}{400--407} (\bibinfo{year}{1995}).

\bibitem{1997ApJ...476L..27W}
\bibinfo{author}{{Wang}, L.}, \bibinfo{author}{{Wheeler}, J.~C.} \&
  \bibinfo{author}{{H{\"o}flich}, P.}
\newblock \bibinfo{title}{{Polarimetry of the Type IA Supernova SN 1996X}}.
\newblock \emph{\bibinfo{journal}{\apjl}} \textbf{\bibinfo{volume}{476}},
  \bibinfo{pages}{L27--L30} (\bibinfo{year}{1997}).
\newblock \eprint{astro-ph/9609178}.

\bibitem{1991A&A...246..481H}
\bibinfo{author}{{Hoflich}, P.}
\newblock \bibinfo{title}{{Asphericity Effects in Scattering Dominated
  Photospheres}}.
\newblock \emph{\bibinfo{journal}{\aap}} \textbf{\bibinfo{volume}{246}},
  \bibinfo{pages}{481} (\bibinfo{year}{1991}).

\bibitem{2015ApJ...813....2M}
\bibinfo{author}{{Martin}, D.} \emph{et~al.}
\newblock \bibinfo{title}{{Neutrino-driven Winds in the Aftermath of a Neutron
  Star Merger: Nucleosynthesis and Electromagnetic Transients}}.
\newblock \emph{\bibinfo{journal}{\apj}} \textbf{\bibinfo{volume}{813}},
  \bibinfo{pages}{2} (\bibinfo{year}{2015}).
\newblock \eprint{1506.05048}.

\bibitem{2017PhRvD..96l4005B}
\bibinfo{author}{{Bovard}, L.} \emph{et~al.}
\newblock \bibinfo{title}{{r -process nucleosynthesis from matter ejected in
  binary neutron star mergers}}.
\newblock \emph{\bibinfo{journal}{\prd}} \textbf{\bibinfo{volume}{96}},
  \bibinfo{pages}{124005} (\bibinfo{year}{2017}).
\newblock \eprint{1709.09630}.

\bibitem{1994A&A...289..492H}
\bibinfo{author}{{Hillier}, D.~J.}
\newblock \bibinfo{title}{{The calculation of continuum polarization due to the
  Rayleigh scattering phase matrix in multi-scattering axisymmetric
  envelopes}}.
\newblock \emph{\bibinfo{journal}{\aap}} \textbf{\bibinfo{volume}{289}},
  \bibinfo{pages}{492--504} (\bibinfo{year}{1994}).

\bibitem{1996ApJ...461..847W}
\bibinfo{author}{{Wood}, K.}, \bibinfo{author}{{Bjorkman}, J.~E.},
  \bibinfo{author}{{Whitney}, B.} \& \bibinfo{author}{{Code}, A.}
\newblock \bibinfo{title}{{The Effect of Multiple Scattering on the
  Polarization from Axisymmetric Circumstellar Envelopes. II. Thomson
  Scattering in the Presence of Absorptive Opacity Sources}}.
\newblock \emph{\bibinfo{journal}{\apj}} \textbf{\bibinfo{volume}{461}},
  \bibinfo{pages}{847} (\bibinfo{year}{1996}).

\bibitem{2003ApJ...593..788K}
\bibinfo{author}{{Kasen}, D.} \emph{et~al.}
\newblock \bibinfo{title}{{Analysis of the Flux and Polarization Spectra of the
  Type Ia Supernova SN 2001el: Exploring the Geometry of the High-Velocity
  Ejecta}}.
\newblock \emph{\bibinfo{journal}{\apj}} \textbf{\bibinfo{volume}{593}},
  \bibinfo{pages}{788--808} (\bibinfo{year}{2003}).
\newblock \eprint{astro-ph/0301312}.

\bibitem{2011MNRAS.415.3497D}
\bibinfo{author}{{Dessart}, L.} \& \bibinfo{author}{{Hillier}, D.~J.}
\newblock \bibinfo{title}{{Synthetic line and continuum linear-polarization
  signatures of axisymmetric Type II supernova ejecta}}.
\newblock \emph{\bibinfo{journal}{\mnras}} \textbf{\bibinfo{volume}{415}},
  \bibinfo{pages}{3497--3519} (\bibinfo{year}{2011}).
\newblock \eprint{1104.5346}.

\bibitem{2004ApJ...610..876K}
\bibinfo{author}{{Kasen}, D.}, \bibinfo{author}{{Nugent}, P.},
  \bibinfo{author}{{Thomas}, R.~C.} \& \bibinfo{author}{{Wang}, L.}
\newblock \bibinfo{title}{{Could There Be a Hole in Type Ia Supernovae?}}
\newblock \emph{\bibinfo{journal}{\apj}} \textbf{\bibinfo{volume}{610}},
  \bibinfo{pages}{876--887} (\bibinfo{year}{2004}).
\newblock \eprint{astro-ph/0311009}.

\bibitem{2016MNRAS.462.1039B}
\bibinfo{author}{{Bulla}, M.} \emph{et~al.}
\newblock \bibinfo{title}{{Predicting polarization signatures for
  double-detonation and delayed-detonation models of Type Ia supernovae}}.
\newblock \emph{\bibinfo{journal}{\mnras}} \textbf{\bibinfo{volume}{462}},
  \bibinfo{pages}{1039--1056} (\bibinfo{year}{2016}).
\newblock \eprint{1607.04081}.

\bibitem{Plaszczynskiet2014}
\bibinfo{author}{{Plaszczynski}, S.}, \bibinfo{author}{{Montier}, L.},
  \bibinfo{author}{{Levrier}, F.} \& \bibinfo{author}{{Tristram}, M.}
\newblock \bibinfo{title}{{A novel estimator of the polarization amplitude from
  normally distributed Stokes parameters}}.
\newblock \emph{\bibinfo{journal}{\mnras}} \textbf{\bibinfo{volume}{439}},
  \bibinfo{pages}{4048--4056} (\bibinfo{year}{2014}).
\newblock \eprint{1312.0437}.

\end{thebibliography}


\begin{addendum}
 \item \revised{The authors thank the anonymous referees for their constructive feedback which
improved this paper}. M.B. acknowledges support from the Swedish Research Council (Vetenskapsr\aa det) and the Swedish National Space Board. S.C. acknowledges support from ASI grant I/004/11/3.  K.K. is supported by Japanese Society for the Promotion of Science (JSPS) Kakenhi Grant-in-Aid for Scientific Research (No.~JP16H06342,
No.~JP17H01131, No.~JP18H04595). J.R.M. is supported through a Royal Society University Research Fellowship. K.T. is supported by JSPS Kakenhi No. 15H05437 and No. 18H01245, and also by a JST grant Buidling of Consortia for the Development of Human Resources in Science and Technology. J.B. is supported by a University of Sheffield PhD studentship.
\item[Author contributions] All authors contributed to the work presented in this paper. M.B. carried out the model simulations and analysis and led the writing of the manuscript. S.C. provided polarimetric data of AT\,2017gfo and helped with the writing. K.K. and M.T. provided theoretical insights on hydrodynamical models, carried out radiative transfer calculations to estimate opacities and helped with the writing of the manuscript.
 \item[Competing Interests] The authors declare that they have no
competing financial interests.
 \item[Correspondence] Correspondence and requests for materials
should be addressed to M.B. or S.C.~(email: mattia.bulla@fysik.su.se, stefano.covino@inaf.it).
\end{addendum}


\begin{figure}
\begin{center}
\includegraphics[width=0.88\textwidth]{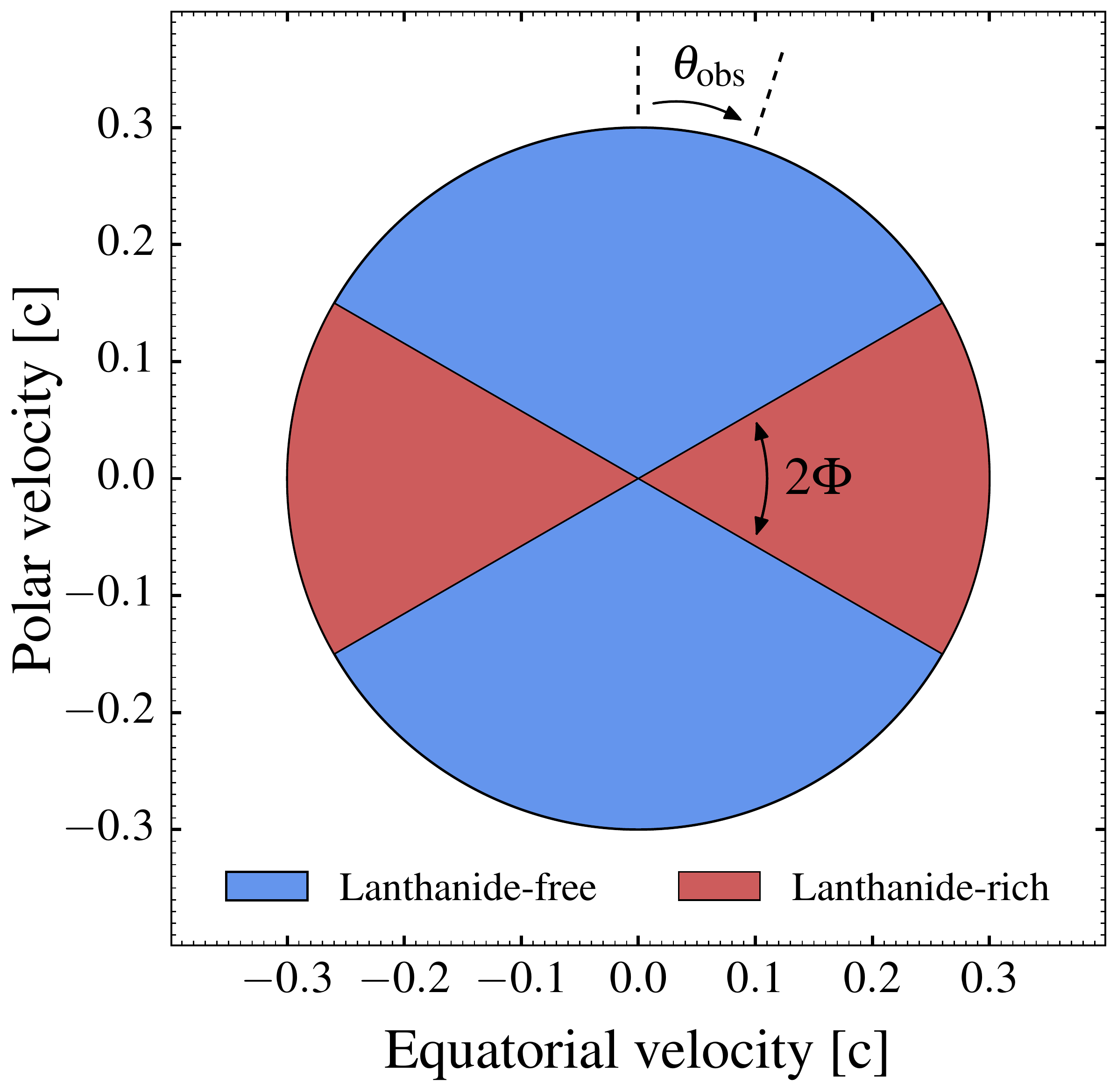}
\caption{\small{Sketch of the fiducial kilonova model used in this work. A meridional cross section of the ejecta is shown. Two ejecta components are adopted: a first component distributed around the equatorial plane and characterized by lanthanide-elements opacities (``lanthanide-rich component", in red) and a second component at higher latitudes characterized by lower opacities (``lanthanide-free component", in blue). These are grossly responsible for the ``red'' and ``blue'' components introduced to model the observed spectra of AT\,2017gfo\cite{Pianetal2017,Smarttetal2017}. The half-opening angle of the lanthanide-rich region is set to $\Phi=30^\circ$. Polarization signals are calculated as a function of viewing angle $\theta_\mathrm{obs}$, where $\cos\theta_\mathrm{obs}=1$ corresponds to the polar direction.}}
\label{fig:sketch}
\end{center}
\end{figure}

\begin{figure}
\begin{center}
\includegraphics[width=0.59\textwidth]{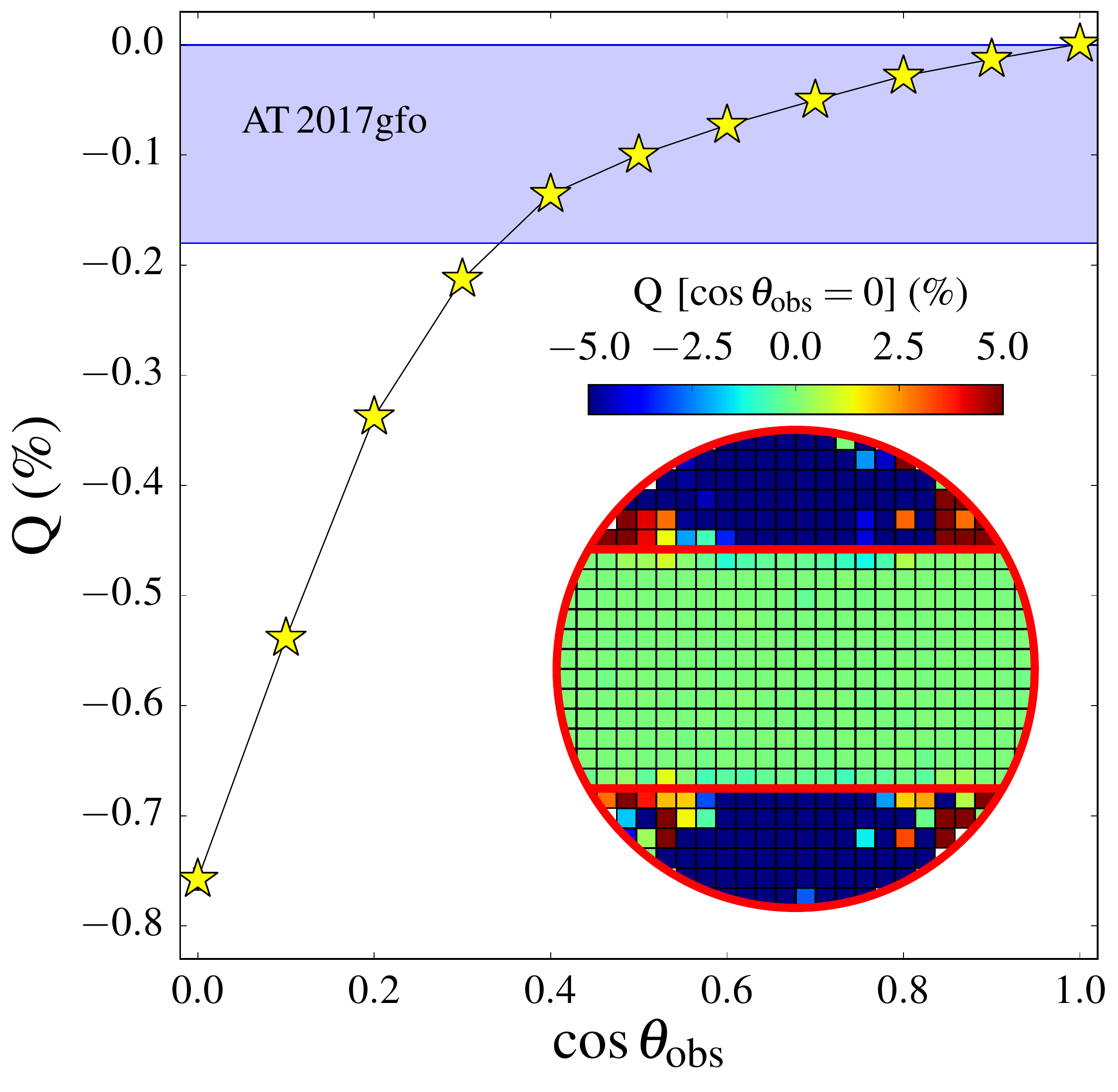}
\caption{\small{Predicted linear polarization at 1.5 days \revised{and at 7000~\AA}~as a function of the viewing angle of the system, $\theta_\mathrm{obs}$ (yellow stars). \revisedtwo{Uncertainties are smaller than the symbol size.} The Stokes parameter $Q$ expresses the difference in intensity between the electric field components along the vertical and horizontal direction. The blue shaded area marks the range of polarization \revisedtwo{estimated} for AT\,2017gfo after removing the interstellar polarization contribution \revisedtwo{(see Methods - Upper limits to AT 2017gfo intrinsic polarization)} from the signal detected in \cite{Covinoetal2017}. The very small level of polarization in AT\,2017gfo is consistent with a system observed within 65 degrees from the polar \revised{axis} ($\cos\theta_\mathrm{obs}\gtrsim0.4$), \revised{a value that is less restrictive but consistent} with independent estimates from the literature\cite{Abbottetal2017a,troja2017,Pianetal2017,Finstadetal2018,Mandel2018}. The inset shows contributions to $Q$ for an equatorial viewing angle ($\cos\theta_\mathrm{obs}=0$), \revised{with the red horizontal lines delimiting the lanthanide-rich ejecta component (cf Fig.~\ref{fig:sketch}) and the size of each pixel equal to 0.025\,c. C}ontributions from regions between the horizontal red lines \revised{are} preferentially depolarized by bound-bound interactions in the lanthanide-rich component (green, $Q\sim0$) while those at higher latitudes \revised{are} preferentially polarized by electron scattering (blue, $Q<0$) in the lanthanide-free component.}}
\label{fig:pol}
\end{center}
\end{figure}

\begin{figure}
\begin{center}
\includegraphics[width=1\textwidth]{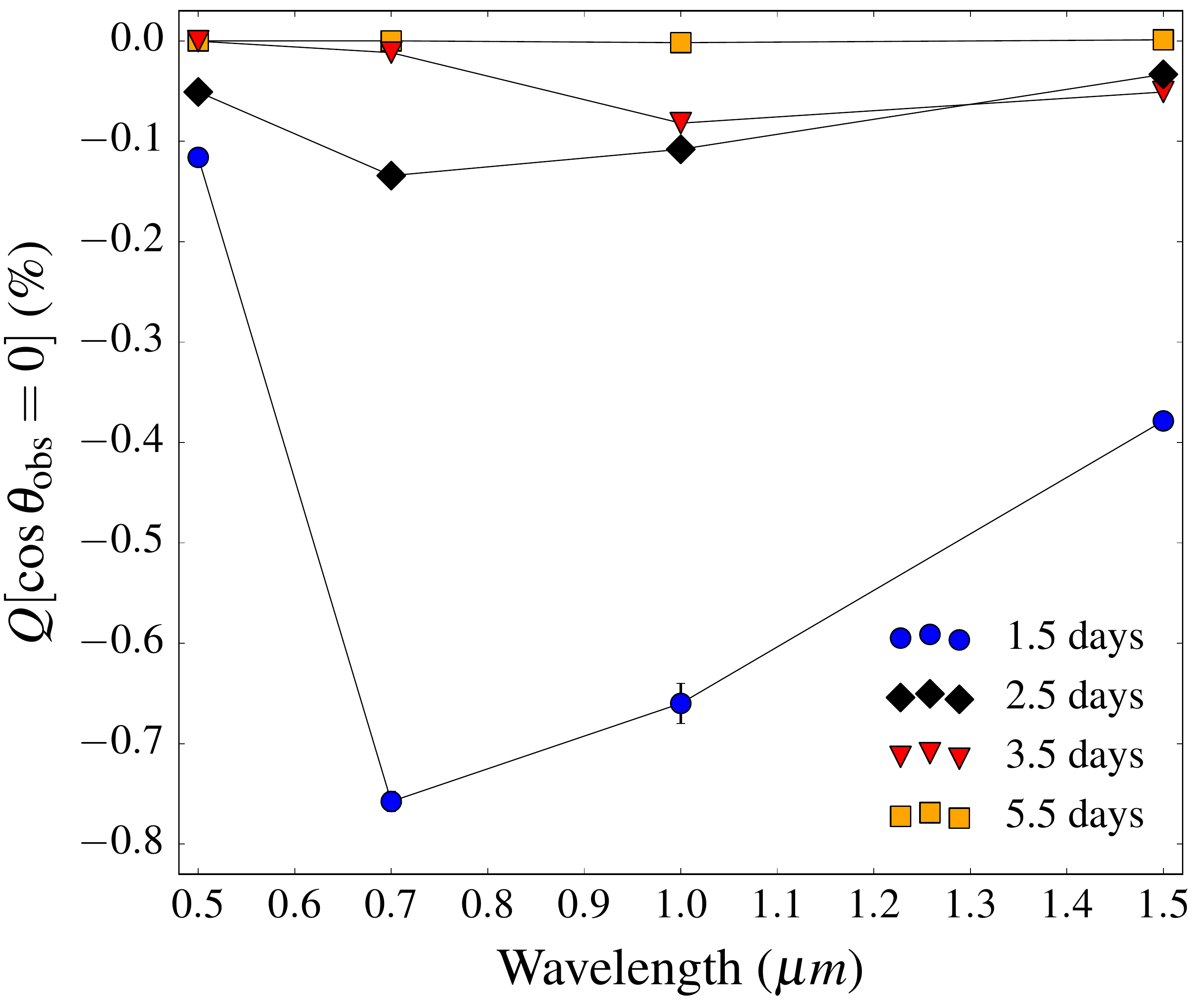}
\caption{\small{Linear polarization $Q$ for an equatorial viewing angle ($\cos\theta_\mathrm{obs}=0$) at different wavelengths and epochs. Wavelengths are chosen to match several often used broadband astronomical filters. The best chances for a positive detection are predicted around 7000~\AA{} (R band) and at relatively early epochs ($1-2$~days from the merger). The polarization in the optical becomes negligible from $2-3$~days after the merger, thus providing a powerful way to characterize the time-independent interstellar polarization in AT\,2017gfo and future kilonova events. Uncertainties are smaller than the symbol size for most datapoints.}}
\label{fig:timewave}
\end{center}
\end{figure}

\begin{figure}
\begin{center}
\includegraphics[width=0.8\textwidth]{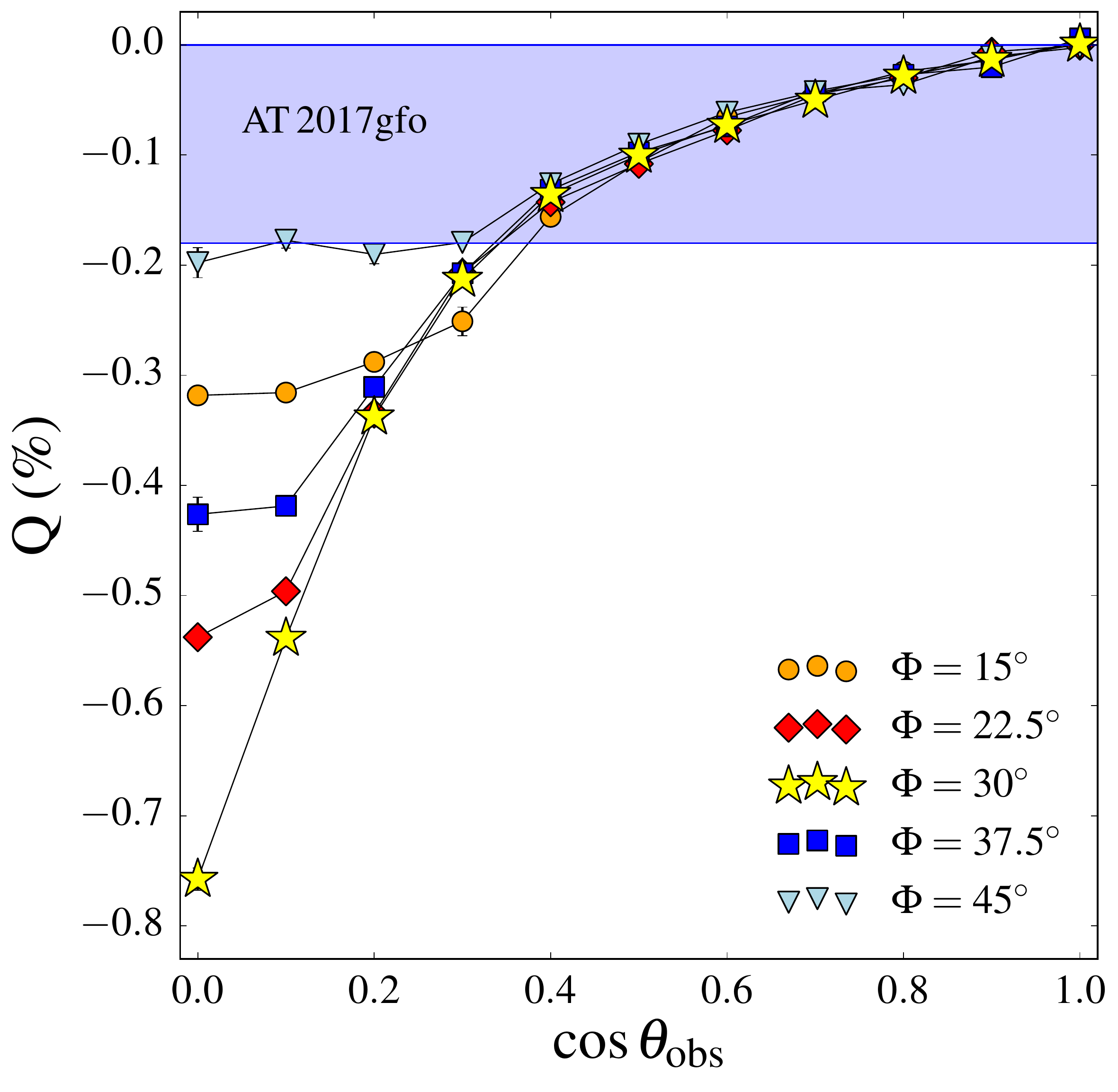}
\caption{\small{Impact of the half-opening angle of the lanthanide-rich ejecta, $\Phi$,  \revised{on the polarization signal predicted at 7000~\AA~and 1.5~days after the merger}. \revisedtwo{The blue shaded area marks the range of polarization estimated for AT\,2017gfo after removing the interstellar polarization contribution (see Methods - Upper limits to AT 2017gfo intrinsic polarization) from the signal detected in \cite{Covinoetal2017}.} At low inclinations ($\cos\theta_\mathrm{obs}\lesssim0.2$), the largest polarization degree is found for $\Phi=30^\circ$, corresponding to the lanthanide-rich ejecta covering half of the full solid angle ($\Delta\Omega=2\pi$). Similar polarization levels are found for viewing angles closer to the polar direction ($\cos\theta_\mathrm{obs}\gtrsim0.4$). The plot highlights how constraints derived in this study for the inclination of AT\,2017gfo, $\cos\theta_\mathrm{obs}\gtrsim0.4$, are rather insensitive  to the choice of $\Phi$.} Uncertainties are smaller than the symbol size for most datapoints.}
\label{fig:opang}
\end{center}
\end{figure}

\begin{figure}
\begin{center}
\includegraphics[width=0.524\textwidth]{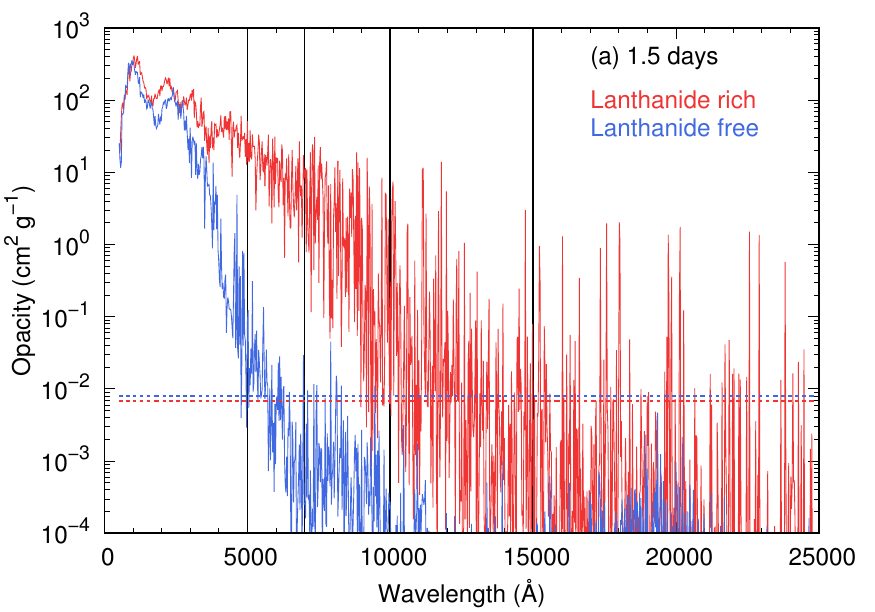}
\includegraphics[width=0.466\textwidth,clip=True,trim=28pt 0pt 0pt 0pt]{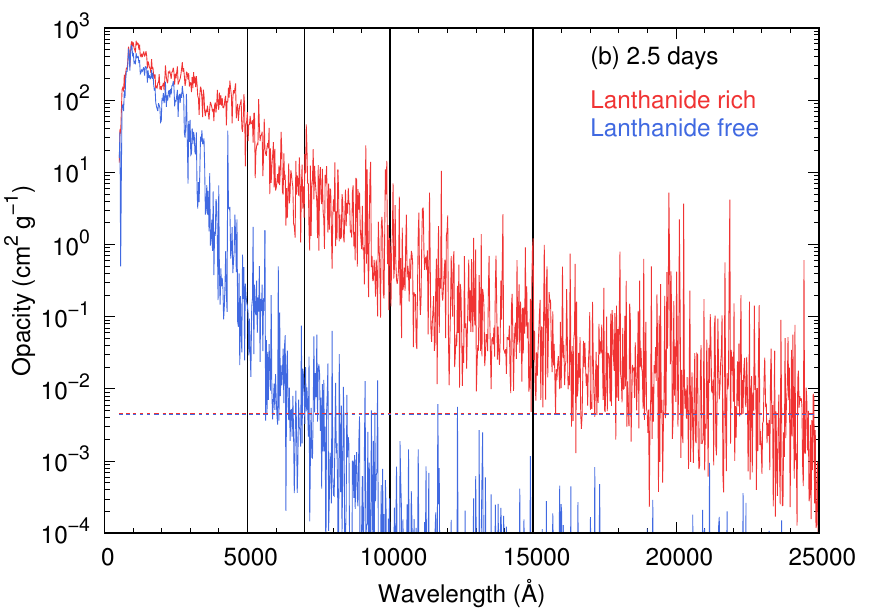}
\includegraphics[width=0.524\textwidth]{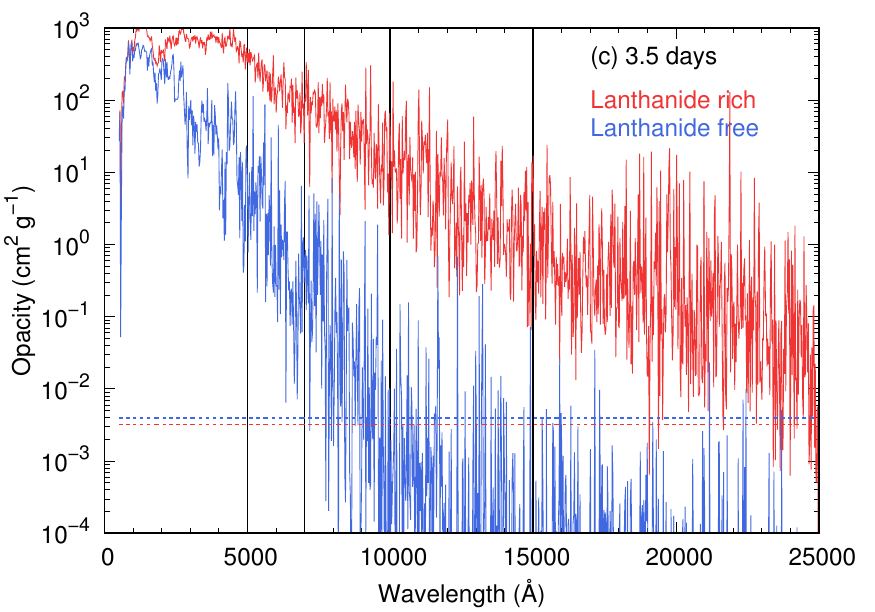}
\includegraphics[width=0.466\textwidth,clip=True,trim=28pt 0pt 0pt 0pt]{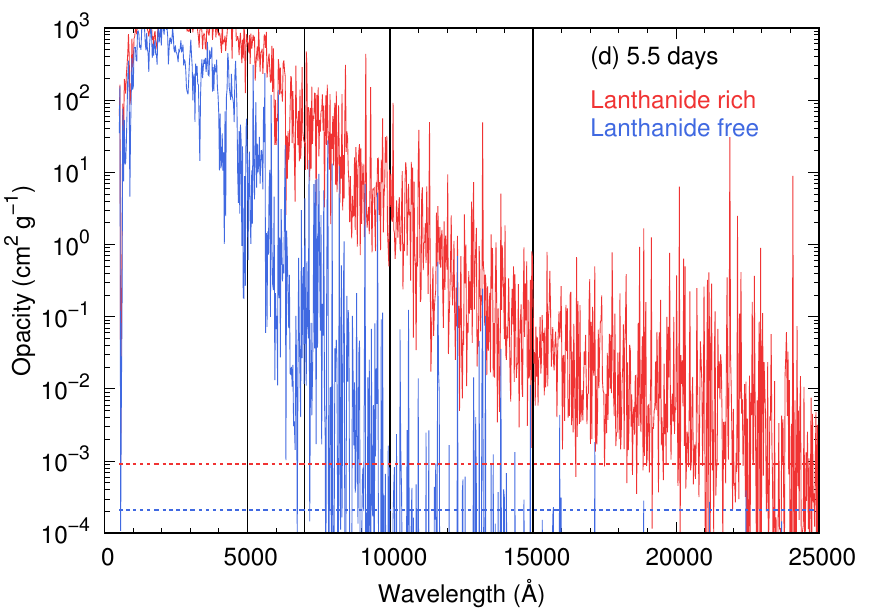}

\caption{\small{Opacities for the lanthanide-free \revised{component at high latitudes} (blue) and lanthanide-rich \revised{component at low latitudes} (red) at 1.5 (panel a), 2.5 (b), 3.5 (c) and 5.5 (d)~days after the merger. Solid lines refer to bound-bound opacities, while horizontal dashed lines to electron scattering opacities. Bound-free and free-free opacities are neglected as they are between $\sim$~5 and 12 orders of magnitudes smaller than electron scattering and bound-bound opacities at epochs and wavelengths (5000, 7000, 10\,000 and 15\,000~\AA, \revised{vertical lines}) considered in this work.}}
\label{fig:opac}
\end{center}
\end{figure}

\begin{figure}
\begin{center}
\includegraphics[width=1\textwidth]{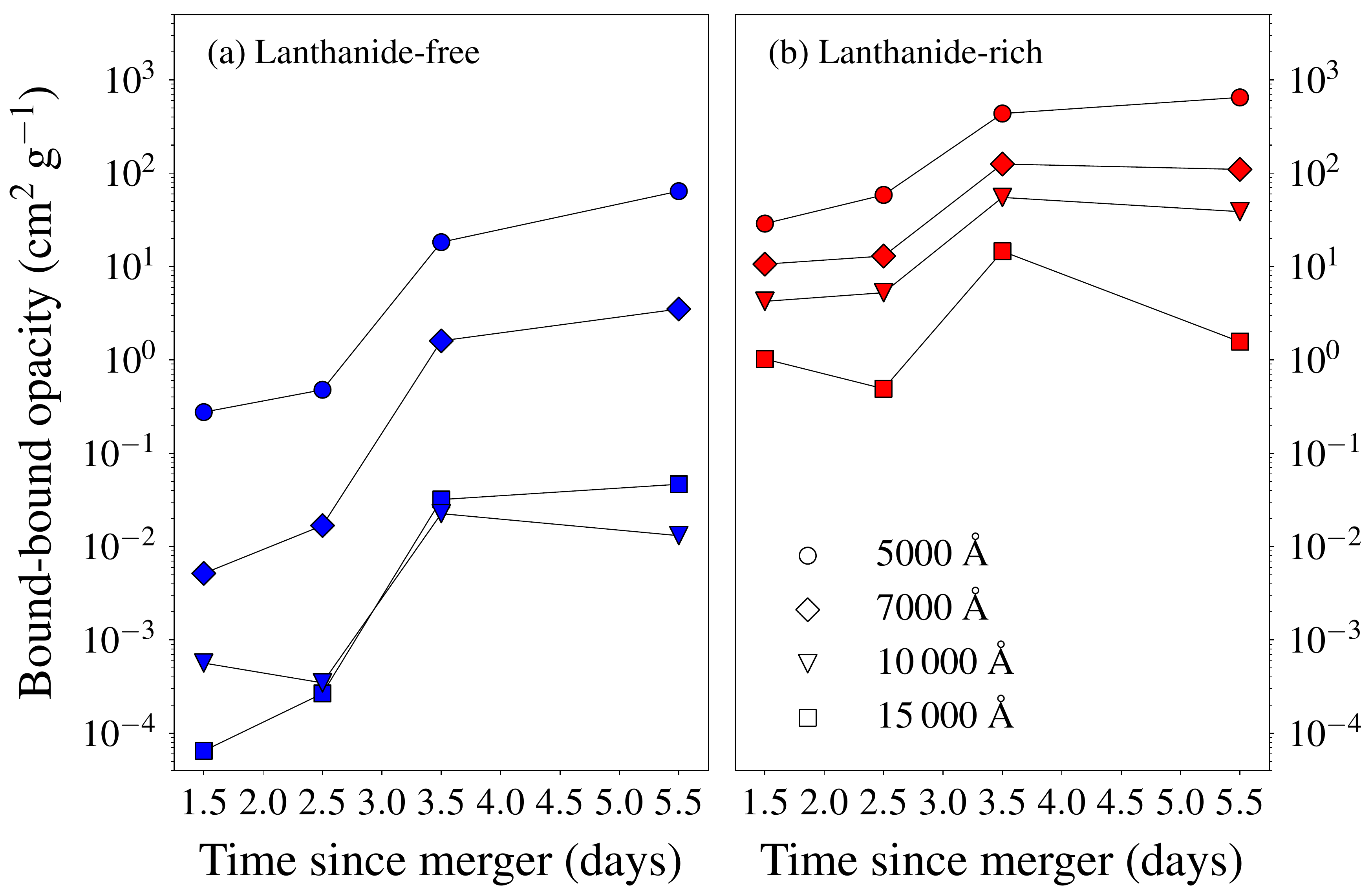}
\caption{\small{\revised{Bound-bound opacities used in the simulations as a function of time since merger. Opacities are shown for lanthanide-free (panel a) and lanthanide-rich (panel b) components at wavelengths (5000, 7000, 10\,000 and 15\,000~\AA) and epochs (1.5, 2.5, 3.5, 5.5~days) considered in this work. Values are calculated by averaging bound-bound opacities of Fig.~\ref{fig:opac} over a range of $\Delta\lambda=500$~\AA~centered at the desired wavelengths.}}}
\label{fig:opac_time_wave}
\end{center}
\end{figure}

\end{document}